\documentclass[aps,prd,10pt,nofootinbib,twocolumn,eqsecnum,showpacs,showkeys,superscriptaddress,preprintnumbers,altaffilletter]{revtex4-1}

\usepackage{graphicx}
\usepackage{amsmath}
\usepackage{multirow}
\usepackage{hyperref}
\usepackage{soul}
\usepackage{graphicx}
\usepackage{dcolumn}
\usepackage{amssymb}
\usepackage{amsmath}
\usepackage{amsfonts}
\usepackage{amsbsy}
\usepackage{color}
\usepackage{rotating}
\usepackage[english]{babel}
\usepackage{multirow}
\usepackage{float}
\usepackage{cleveref}

\DeclareMathOperator\arctanh{arctanh}

\newcolumntype{x}[1]{>{\centering\arraybackslash\hspace{0pt}}p{#1}}

\begin{document}

\title{High accuracy on $H_{0}$ constraints from gravitational wave lensing events}

\date{December 4, 2019}

\author{P. Cremonese}
\email{paolo.cremonese@usz.edu.pl}
\affiliation{Institute of Physics, University of Szczecin, Wielkopolska 15, 70-451 Szczecin, Poland}
\author{V. Salzano}
\email{vincenzo.salzano@usz.edu.pl}
\affiliation{Institute of Physics, University of Szczecin, Wielkopolska 15, 70-451 Szczecin, Poland}


\begin{abstract}
In light of the newly opened and rapidly growing gravitational waves window in multi-messenger astronomy, in order to fully take advantage of the new opportunities we are provided with, new ideas are required for a better and deeper employ of the state-of-the-art probes we handle. Following this goal, here we suggest a method to constrain the cosmological background, and the Hubble constant in particular, by future observations of gravitationally lensed radiation emitted by a single source in both the gravitational wave and the electromagnetic regimes. The lensing of the gravitational wave radiation, in fact, can leave a clear imprinting in the corresponding waveform, and we want to analyze if such kind of measurements can be successfully employed to better constrain the cosmological background. Thus, by making use of wave optics for the gravitational wave lensed signal, and of standard geometrical optics approximation for the electromagnetic one, we study the impact of different cosmological parameters on the value of the arrival time delay due to gravitational lensing, given specific gravitational wave frequencies, mass models of the lens, and redshifts and positions (with respect to the lens) of the source. Although the rate of lensing of gravitational waves is expected to be low, we show that even one single lensing event, combined with a prior on $\Omega_m$ from \textit{Planck}, could provide us with an uncertainty on $H_0$ comparable with present independent probes in a ``pessimistic'' scenario (with a pulsar population similar to present Pulsar Timing Array state), and of two orders smaller in an optimistic one (with a number of observed pulsars as large as that expected from the Square Kilometer Array). Thus, its role in the solution of the Hubble tension could be decisive.
\end{abstract}



\maketitle

\section{Introduction}

After the first combined detection of gravitational and electromagnetic waves \cite{2017PhRvL.119p1101A} from the same source, a new multi-messenger  window has officially opened for astronomy, bringing with it new great opportunities to give much deeper insights and understanding in many of the current main astrophysical \cite{Meszaros:2019xej} and cosmological \cite{Ezquiaga:2018btd,Belgacem:2019pkk} puzzles.

One of the cosmological problems that the multi-messenger GW astronomy could help to solve is the so-called \textit{``Hubble tension''} (see for example \cite{Chang:2019}). Historically speaking, the exact value of the Hubble constant $H_0$, i.e. the rate of expansion of our Universe today, has always been object of intense debates, basically due to its large errors. But in the latest years we have faced a more troublesome scenario: nowadays we are able to perform very precise measurements of $H_0$ in many independent ways, but such determinations do not agree among each other at an alarming statistically significant level.

At the present stage, on one hand we have the latest measurement provided by the \textit{Planck} satellite using the Cosmic Microwave Background (CMB) \cite{PlanckCollab:2018} which gives $H_0=67.36\pm0.54$ km s$^{-1}$ Mpc$^{-1}$. Such a value is generally considered as \textit{inferred}, because it heavily relies on the assumption of a cosmological model (in \textit{Planck} it is a standard $\Lambda$CDM model \cite{Bull:2015stt}), and is strongly connected with \textit{early times} physics (prior to and at recombination). On the other hand, we have local or \textit{late times} observations mainly from the cosmic distance ladder \cite{Riess:2011yx,Riess:2016,Riess:2019cxk}, which have undergone a progressive improvement process, setting the latest estimate at a $1\%$-precision value, $H_0=74.03 \pm 1.42$ km s$^{-1}$ Mpc$^{-1}$. The discrepancy between these two reference measurements is thus established at $ 4.4\sigma$ \textit{``$\ldots$beyond a plausible level of chance''} \cite{Riess:2019cxk}.

Other complementary probes and/or methods have been made available more recently, but without clarifying the picture: upgrades in the analysis of water masers in $NGC4258$ \cite{Reid:2019tiq} have led to $H_0=73.5\pm1.4$ km s$^{-1}$ Mpc$^{-1}$, a $4.2\sigma$ tension with \textit{Planck}; gravitationally-lensed time delays from quasars \cite{Chen:2019,Wong:2019} analyzed within the $H_0$ Lenses in COSMOGRAIL’s Wellspring (H0LiCOW) collaboration \cite{Suyu:2016qxx} have provided $H_0=73.3^{+1.7}_{-1.8}$ km s$^{-1}$ Mpc$^{-1}$, a $2.4\%$ precision measurement with a $3.1\sigma$ tension with \textit{Planck}; the Dark Energy Survey (DES) collaboration \cite{Abbott:2018jhe} has used Type Ia Supernovae and Baryon Acoustic Oscillations in an ``inverse distance ladder'' method to give $H_0=67.8\pm1.3$ km s$^{-1}$ Mpc$^{-1}$ \cite{Macaulay:2018fxi}, which perfectly agrees with \textit{Planck} and is thus in tension with local measurements from distance ladder.

Last but not least, the totally independent estimation provided by GW events is also available now: the LIGO collaboration \cite{Abbott:2017xzu} has obtained $H_0=70.0^{+12.0}_{-8.0}$ km s$^{-1}$ Mpc$^{-1}$ from the multi-messenger detection of a binary neutron star inspiral \citep{2017PhRvL.119p1101A} hence setting in the middle of the contenders, \textit{but} with a much larger error bar than previous probes, thus not helping to solve the tension, at least for now. Improvement have been obtained combining also radio signals \cite{Hotokezaka:2018dfi}, with a final value of $H_0=68.9^{+4.7}_{-4.6}$ km s$^{-1}$ Mpc$^{-1}$, which is, again, due to the relatively large errors, perfectly consistent with both \textit{Planck} and local estimations.

A plethora of solutions have been proposed to explain or solve such a tension. Even if some criticisms inherent to the distance ladder method, both for the procedure and for the existence of a local void \cite{Shanks:2018rka,Shanks:2018dsp} have been raised, they have been replied quite convincingly \cite{Riess:2018kzi,Kenworthy:2019qwq}. Thus, nowadays the main current scenario focuses on translating such observational tension in a tension between our description and understanding of both early and late time physics \cite{Bernal:2016gxb,Verde:2019ivm,Jimenez:2019}.

While waiting for future surveys being operative and, possibly, decisive in setting clarity on this topic \cite{Bengaly:2019oxx}, the main goal now should be to find new potential probes (or new alternative ways to employ current probes) which might add information to the debate and be competitive with present measurements for what concerns the achievable precision. Following this main objective, in this work we will study which kind of information may be inferred from the measurement and the analysis of the arrival time difference (ATD) from a gravitational lensing event\footnote{In this paper by "gravitational lensing" we refer to strong gravitational lensing.} of both GWs and EM signals in a multi-messenger detection.

In \cite{Takahashi:2003,Takahashi:2016jom,Cremonese:2018} it was studied which kind of ATD one should expect to measure by varying mass modelling of the lens, the (redshift) position of the source, and the relative angular position between the lens and the source. It was found that a significant ATD is expected when considering a super massive binary black hole (SMBBH) emitting both EM radiation and GWs, with a very small frequency ($f \sim 10^{-8}$ Hz), lensed by a galaxy with a mass $M_{lens} \sim 10^{11}~M_{\odot}$. Although the probability to observe such event(s) is of the order of $10^{-3}-10^{-4}$ \cite{Takahashi:2016jom}, it has been shown that they can modify with a clear signature the waveform of a detected GW. Thus, it would be interesting to investigate which kind of additional information, on the cosmological side, one could retain from at least one event.

Here we will explore how this imprinting might be used for cosmological purposes to infer the value of $H_0$ and, more crucially, with which precision. As for GWs signal, we will assume they are detected and measured by pulsar timing array (PTA) observatories, like the presently running International PTA (IPTA), the collaboration between the three major PTA collaborations \cite{2016IPTA} and, for the next future, by the Square Kilometer Array (SKA) \cite{Lazio:2013,Pulido:2015cla,Bull:2018lat}.

The article is organized as follows: in Sec.~\ref{time delay} we explain how to approach an ATD analysis in both GW and EM gravitational lensing events, and how it depends on the cosmological parameters; in Sec.~\ref{Methodology} we explain how we proceed to obtain the uncertainty on the $H_0$ measurement; in Sec.~\ref{results}, we summarize our results and discuss them; conclusions are drawn in Sec.~\ref{conclusions}.

\section{Arrival time difference}\label{time delay}

\begin{figure}[htbp]
\centering
\includegraphics[width=0.35\textwidth]{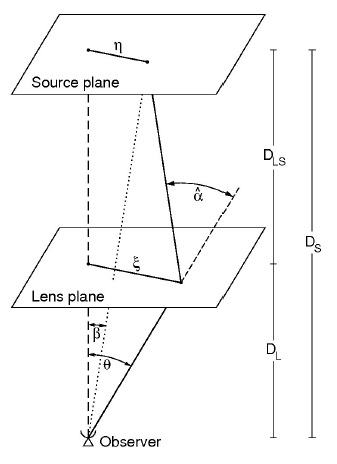}
\caption{Geometry of a gravitational lens system: $\beta$ is the angular position of the source; $\theta$ is the angular position of the image; $\hat{\alpha}$ is the deflection angle. The angular diameter distances between observer and lens, lens and source, and observer and source are $D_{L}$, $D_{LS}$, and $D_{S}$, respectively. Figure from \cite{gralen.boo}.}\label{lens}
\end{figure}

The basic core of our study is the possibility to detect a gravitational lensing event by some foreground object (lens) which deviates both the EM and GW counterparts from the source. The typical depiction of a gravitational lensing system is shown in \figurename~\ref{lens}. A difference in the corresponding (gravitational lensing-)time delays will depend on the relation between the wavelength of the lensed radiation and the mass of the lens, once all other possible sources of delays have been excluded and/or deleted (e.g. delays generating at the source). In fact, while for the EM signal this relation is typically negligible, i.e. the EM gravitational lensing can be described and studied in the \textit{geometrical optics} regime, for GW we might need a \textit{wave optics} approach, if some conditions are verified. In particular, when the GW has an extremely low frequency ($f\sim 10^{-8}$ Hz, i.e. very large wavelength, $\lambda \sim 1$ parsec) and the lensing object is a galaxy ($M_{\rm{galaxy}} \simeq 10^{11} M_\odot$), wave optics must be used to study the GW lensed signal and the geometrical optics approximation generally used for the EM counterpart must be abandoned.

The boundary between the two approximations is defined, among others, in \cite{Takahashi:2016jom}: geometrical optics approximation breaks when
\begin{equation}
M\leq10^5 M_{\odot}\left(\frac{f}{\text{Hz}}\right)^{-1},
\end{equation}
where $f$ is the frequency of the lensed radiation and $M$ is the mass of the lens. For a galaxy with $M\sim10^{11}\ M_\odot$, if $f \lesssim 10^{-6}$ Hz (or $\lambda\gtrsim 1.5\cdot10^{14}$ m), wave optics should be used. For $f \approx10^{-8}$ Hz, a typical SMBBH GW frequency probed by PTA, wave optics must be used for lenses with $M \leq 10^{13}\ M_\odot$.

When this should happen, we might have the possibility to detect an ATD in the time delays of both EM and GW signals. Such ATD is operationally defined as
\begin{equation}\label{arrtdiff}
\Delta t_{\rm EM-GW}(y,w)=t_{\rm EM}(y)-t_{\rm GW}(y,w),
\end{equation}
where $t_{\rm EM}$\footnote{Note that for completeness one should write $t_{\rm EM,\pm}$, where $\pm$ refers to the different magnification of the images. Here, we will focus only on the brighter image.} and $t_{\rm GW}$ are the gravitational lensing time delays of the EM and GW signal, respectively; $y$ is the position of the source in units of a characteristic radius on the lens plane, and $w$ is the dimensionless frequency of the GW. See \figurename~\ref{lens} for geometrical configuration and next pages for definitions.

In order to understand how the ATD depends on the cosmological parameters, let us see how it is defined. In geometrical optics, the time delay is
\begin{equation}\label{timedelGO}
    t(\boldsymbol{\theta},\boldsymbol{\beta})=\frac{1+z_L}{c}\frac{D_L D_S}{D_{LS}}\left[\frac{1}{2}(\boldsymbol{\theta}-\boldsymbol{\beta})^2-
    \hat{\Psi}(\boldsymbol{\theta})\right] ,
\end{equation}
where angles and distances are defined in \figurename~\ref{lens}; $z_L$ is the redshift of the lens; and $\hat{\Psi}(\boldsymbol{\theta})$ is the effective lens potential defined as
\begin{equation}\label{lenspot}
\hat{\Psi}(\boldsymbol{\theta}) = \frac{D_{LS}}{D_L D_S}\frac{2}{c^2}\int \Phi(D_L\boldsymbol{\theta},\mathrm{z})d\mathrm{z}\, ,
\end{equation}
where $\Phi$ is the gravitational potential of the lens and $\mathrm{z}$ is the line-of-sight coordinate. An alternative way to write the previous expression in term of dimensionless quantities is:
\begin{equation}\label{timedelNODim}
t(\boldsymbol{x},\boldsymbol{y}) = \frac{1+z_L}{c} \frac{D_S \xi^{2}_0}{D_L D_{LS}} \left[ \frac{1}{2} \left( \boldsymbol{x}-\boldsymbol{y} \right)^2 - \Psi(\boldsymbol{x})\right]\; ,
\end{equation}
where $\boldsymbol{\xi}_0$ is a reference scale length on the lens plane whose value depends on the mass model of the lens; $\boldsymbol{x}=D_L\boldsymbol{\theta}/\xi_0$ and $\boldsymbol{y}=D_L\boldsymbol{\eta}/D_S \boldsymbol{\xi}_0=D_L\boldsymbol{\beta}/\boldsymbol{\xi}_0$ are the dimensionless relative positions of the source and of the image; and $\Psi(\boldsymbol{x}) = D^{2}_l/\xi^{2}_{0}\hat{\Psi}(\boldsymbol{\theta})$ is the dimensionless lens potential. Finally, we define the dimensionless time delay as
\begin{equation}\label{eq:tGOdimless}
T(\boldsymbol{x}, \boldsymbol{y})=\frac{c}{1+z_L}\frac{D_L D_{LS}}{D_S \xi^{2}_0}\cdot t(\boldsymbol{x},\boldsymbol{y})\, .
\end{equation}
In the case of an EM signal, we are in the geometrical optics regime, and we simply have:
\begin{equation}\label{eq:tEMdimless}
T_{\rm EM}(\boldsymbol{x}, \boldsymbol{y}) = T(\boldsymbol{x}, \boldsymbol{y})\, .
\end{equation}
On the other hand, in the wave optics regime, the time delay is defined \cite{Takahashi:2016jom} as
\begin{equation}\label{timedelWO}
T_{GW}(w,\boldsymbol{y})\equiv -\frac{i}{w}\ln{\left(\frac{F(w,\boldsymbol{y})}{|F(w,\boldsymbol{y})|}\right)} ,
\end{equation}
where $F(w,\boldsymbol{y})$ is the amplification factor \cite{gralen.boo}
\begin{equation}\label{ampfactor}
F(w,\boldsymbol{y})=\frac{w}{2\pi i}\int \mathrm{d}^2x\exp[iwT(\boldsymbol{x},\boldsymbol{y})]\; ,
\end{equation}
where the dimensionless frequency of the GW radiation is $w=\frac{1+z_L}{c}\frac{D_S \xi^{2}_{0}}{D_LD_{LS}} 2\pi f$, with $f$ the dimensional frequency.

\section{Methodology}\label{Methodology}

In order to check with which precision the ATD can measure some cosmological parameters, and $H_0$ in particular, we will proceed as follows:
\begin{itemize}
 \item we calculate $\Delta t_{\rm{EM-GW}}$, from Eqs.~\eqref{arrtdiff}~-~\eqref{timedelNODim}~-~\eqref{eq:tEMdimless} and \eqref{timedelWO}, for a large set of input vectors built on the cosmological parameters $\{\Omega_m, H_0, w_0\}$. In particular\footnote{The format is $variable :=$ [start point, end point, step].}: $\Omega_m:= [0.20,\ 0.40,\ 0.01]$, $H_0:= [60.0,\ 80.0,\ 0.1]$ and $w_0:= [-3.50,\ 1.50,\ 0.01]$. We choose the range intervals for the parameters wide enough in order to have a full view of the behaviour of the ATD in the cosmological space;
 \item for each observational scenario (as described in Sec.~\ref{sec:sensitivity}) we calculate the corresponding expected uncertainty\footnote{Just for clarity of notation we will define the uncertainty on the EM-GW ATD as $\sigma_{\Delta t} \equiv \sigma_{\Delta t_{\rm EM-GW}}$.}, $\sigma_{\Delta t}$;
 \item we assume an independent prior on $\Omega_m$ from \textit{Planck}\footnote{\url{https://wiki.cosmos.esa.int/planck-legacy-archive/index.php/Cosmological_Parameters}}, $\Omega_m = 0.3061\pm 0.0052$. Even though this value assumes a $\Lambda$CDM cosmology, it does not affect the generality of the final result. Indeed, for example, choosing a value of $\Omega_m=0.2968 \pm 0.0069$, taken from \textit{Planck}, assuming a $w$CDM cosmology, does not change the results;
 \item we infer the uncertainty on $H_0$ by crossing the above prior with the range in the ATD spanned by assuming the uncertainty $\sigma_{\Delta t}$.
\end{itemize}
In the following subsections we will provide more details about each step.

\subsection{Lens models}\label{sec:lensmodel}

In order to calculate the ATD $\Delta t_{\rm{EM-GW}}$, the first ingredient we need to supply is a mass model for the lens. Many examples are available in the literature about the use of wave optics in gravitational lensing, examining especially the simplest case of a point mass lens \cite{gralen.boo, gralen.art}. But as models get more complicated, the number of papers diminishes and only a few can be found with an in-depth calculations of all the useful expressions. That is because, while for a point mass lens the amplification factor integral is analytically solvable, in the case of more realistic models like the Singular Isothermal Sphere (SIS), or the Navarro-Frenk-White (NFW) one \cite{Navarro:1995iw,1999PThPS.133..137N, Takahashi:2003, Meneghetti}, we can only refer to numerical solutions.

We have chosen to focus our results on two mass models for the lens, exactly SIS and NFW. In order to calculate the ATD, we first have to solve the lens equation \cite{gralen.boo},
\begin{equation}\label{eq:lenseq}
\boldsymbol{\beta} = \boldsymbol{\theta} - \boldsymbol{\alpha}(\boldsymbol{\theta})\;,
\end{equation}
where the reduced deflection angle is defined as $\boldsymbol{\alpha}(\boldsymbol{\theta}) = D_{LS}/D_{S} \hat{\boldsymbol{\alpha}}(\boldsymbol{\theta})$, with $\hat{\boldsymbol{\alpha}}$ the deflection angle. Equivalently, in dimensionless quantities the lens equation can be written as
\begin{equation}\label{eq:lenseq_dimless}
\boldsymbol{y} = \boldsymbol{x} - \boldsymbol{\alpha}(\boldsymbol{x}),
\end{equation}
where $\boldsymbol{\alpha}(\boldsymbol{x}) = D_L D_{LS} / \xi_0 D_S \hat{\boldsymbol{\alpha}}(\xi_0 \boldsymbol{x})$ is the scaled deflection angle \cite{Meneghetti} related to the dimensionless lens potential by
\begin{equation}\label{eq:alpha}
\nabla_{\boldsymbol{x}} \Psi(\boldsymbol{x}) = \boldsymbol{\alpha}(\boldsymbol{x})\; .
\end{equation}
Actually, we will solve Eq.~\eqref{eq:lenseq_dimless} in order to have $\boldsymbol{x}(\boldsymbol{y})$, i.e. our main independent variable will be the position of the source on the lens plane, $\boldsymbol{y}$ (that is why in Eq.~\eqref{arrtdiff} we express the ATD only as function of $y$ and $w$).

\subsubsection{Singular Isothermal Sphere}\label{par:SIS}

The mass density profile of a SIS lens is given by:
\begin{equation}
\rho(r)=\frac{\sigma_\ast^2}{2\pi G}\frac{1}{r^2}~,
\end{equation}
where $\sigma_\ast$ is the velocity dispersion of the stars in the galaxy and $r$ the distance from the center of the galaxy. Assuming an axially symmetric lens, the lens potential, Eq.~\eqref{lenspot}, becomes
\begin{equation}
\hat{\Psi}(\theta) = \frac{4\pi \sigma_\ast^2}{c^2} \frac{D_{LS}}{D_S}\theta\; ,
\end{equation}
or, in dimensionless quantities,
\begin{equation}\label{eq:potSIS}
\Psi(x) = x \; ,
\end{equation}
where the characteristic radius in this case, is $\xi_0 = D_L \theta_{E}$, with $\theta_{E}$ the Einstein radius obtained from Eq.~\eqref{eq:lenseq} when $\beta = 0$. To calculate the time delay in the wave optics regime, we have to obtain first the amplification factor. Assuming spherical symmetry for the lens, and using dimensionless quantities, Eq.~\eqref{ampfactor} becomes \cite{1999PThPS.133..137N}
\begin{eqnarray} \label{eq:besselint}
F(w,y) &=& -iwe^{iwy^2/2} \\
&\times& \int_0^\infty \mathrm{d}x\,x\,J_0(wxy)\exp\left\{ iw\left[ \frac{1}{2}x^2-\Psi(x)\right]\right\}\,, \nonumber
\end{eqnarray}
where $J_0$ is the Bessel function of zeroth order. This integral does not have an analytic solution, therefore we must compute it numerically.

Finally, as this model is fully characterized by the parameter $\sigma_{\ast}$, we will consider a standard stellar dispersion value, $\sigma_{\ast} = 220$ km/s.

\subsubsection{Navarro-Frenk-White}\label{par:NFW}

The density profile in this case is \cite{Navarro:1995iw}
\begin{equation}
\rho(r)=\frac{\rho_s}{\frac{r}{r_s}\left(1+\frac{r}{r_s}\right)^2}~,
\end{equation}
where $\rho_s$ is the characteristic density of the halo and $r_s$ is its scale radius (it will be also considered as the characteristic length on the lens plane, $\xi_0 $). The dimensionless lens potential is
\begin{equation}\label{eq:potNFW}
\Psi(x) = \frac{16\pi G}{c^2} \rho_s r_s \frac{D_LD_{LS}}{D_S} g(x)\; ,
\end{equation}
with the function $g(x)$ given by
\begin{eqnarray}\label{eq:func_g}
g(x) &=& \frac{1}{2} \left\{\left[2 \log (x)-\log \left(-4 \sqrt{x^2-1}+4 i\right)\right]\right. \nonumber \\
     &\times& \left.\Bigl[\log \left(-\sqrt{x^2-1}+i\right)-i \arctan\left(\sqrt{x^2-1}\right)\Bigr]\right. \nonumber \\
     &+& \left. i \log \left(\sqrt{x^2-1}+i\right) \arctan\left(\sqrt{x^2-1}\right)\right\} \nonumber \\
     &-&\frac{1}{8} (\pi -2 i \log 2)^2\, .
\end{eqnarray}
For Eq.~\eqref{eq:func_g}, we are using here the same name $g(x)$ as in \cite{Meneghetti} to ease the comparison between our Eq.~\eqref{eq:potNFW} and its standard expression which makes use of the functions $\arctan$ and $\arctanh$. Our formula is completely equivalent to those ones but more helpful from the numerical point of view because there is no need to switch between different definitions of $g(x)$ depending on the range of $x$ (see for example Pag.~37 of \cite{Meneghetti}), thus leading to faster calculations.

For the NFW case, we will analyze a realistic galaxy model as observed and described in \cite{Buote:2019}.

\subsection{Cosmological dependencies}

At this point, it is interesting to have an insight on how the ATD might depend on the cosmological parameters, and in particular on $H_0$. First of all, one should note that most of the dependence comes from the angular diameter distance terms, which are defined as
\begin{equation}\label{eq:ang_dist}
D_A(z) = \frac{1}{1+z} \int^{z}_{0} \frac{c\, dz'}{H(z')}\, .
\end{equation}
Thus, the angular diameter distances depend on the cosmological background through the Hubble parameter. In this paper we have considered two different cosmological scenarios: a standard $\Lambda$CDM model and a quiessence scenario, whose expansion histories can be described by the following expression:
\begin{eqnarray}\label{hubbleparam}
H^2(z) &=& H_0^2\,[\Omega_k (1+z)^2 +\Omega_m(1+z)^3 \nonumber \\
    &+& \Omega_{\rm DE}(1+z)^{3(1+w_0)}]\; ,
\end{eqnarray}
with: $\Omega_m$, the matter density parameter today; $\Omega_k$, the spatial curvature parameter; $\Omega_{DE} = 1 - \Omega_m - \Omega_k$, the dark energy density parameter; and $w_0$ the dark energy equation of state parameter, which will be $-1$, in the case of dark energy as a cosmological constant, and $ const. \neq -1$ in the quiessence one. Our fiducial cosmological model will have $\Omega_k=0$ and $\Omega_m = 0.3061$ as from latest \textit{Planck} results.

From Eq.~(\ref{eq:ang_dist}) one can easily derive the dependence on $H_0$ of the multiplicative factor appearing in Eq.~(\ref{timedelGO}): the ratio $D_L D_S/D_{LS}$ is simply $\propto H^{-1}_{0}$. For other cosmological parameters, like $\Omega_m$ and $w_0$ the dependence is less trivial, because of the required integrals which are also on different redshift intervals (depending on lens and source positions).

The final dependence of the ATD on $H_0$ is also connected to the mass model employed for the lens. In fact, for SIS, we can see how the EM time delay $t_{EM} \propto H^{-1}_{0}$, while the GW time delay $t_{GW}$ has not such an easy dependence because of the integral Eq.~(\ref{eq:besselint}) which is employed in its derivation (although one can easily recognize that $w \propto H^{-1}_{0}$). For the NFW case, both $t_{EM}$ and $t_{GW}$ do not scale so easily with $H_0$, because of the non-linear dependence of the lens equation Eq.~(\ref{eq:lenseq_dimless}) on $H_0$ in the former case, and also because of the integral in the amplification factor.

For all these reasons, we have mainly relied on numerical analysis, using our own \textsc{Mathematica} code to calculate the ATD and find out its cosmological dependencies, and our own \textsc{Python} code for data analysis.

\subsection{Sensitivity of observations}
\label{sec:sensitivity}

A crucial ingredient in this work is the uncertainty on the GW time delay. In \cite{Cutler:1994} and \cite{Takahashi:2016jom} it is shown that, in a matched filtering analysis, the phase of the waveform can be measured with an accuracy corresponding to the inverse signal-to-noise ratio, $\sigma\approx(S/N)^{-1}$. For example, if $S/N=10$, we can measure the phase difference if $\omega\Delta T_{\rm EM-GW}\gtrsim 10^{-1}$ rad. From \cite{Huerta:2015}, the signal-to-noise ratio $\rho^2$ is defined as
\begin{equation}
\rho^2=\hat{\rho}^2\cdot (1+z)^4\left(\frac{f_{\rm orb}}{f_{\rm obs}}\right)^{-2/3},
\label{snreq}
\end{equation}
where $f_{\rm orb}$ is the orbital frequency of the SMBBH emitting the radiation, $f_{\rm obs}$ is the lowest frequency detectable by the PTAs, and with
\begin{eqnarray}\label{rhohat}
\hat{\rho}^2 &=& 4.26\cdot10^{-2}N_{\rm p}(N_{\rm p}-1)\left(\frac{\mathcal{M}}{10^8 M_\odot}\right)^{10/3} \left(\frac{T_{\rm obs}}{10\text{ yr}}\right)^{5/3} \nonumber \\
&\times& \left(\frac{100\text{ Mpc}}{d_{\rm L}}\right)^2\left(\frac{100\text{ ns}}{\sigma_{\rm rms}}\right)\left(\frac{0.05\text{ yr}}{\Delta \tau}\right),
\end{eqnarray}
where $N_{\rm p}$ is the number of pulsars in the PTA, $\mathcal{M}=\frac{(M_1M_2)^{3/5}}{(M_1+M_2)^{1/5}}$ is the chirp mass of the SMBBH system, $T_{\rm obs}$ is the total baseline time of observation, $d_{\rm L}$ the luminosity distance to the source, $\sigma_{\rm rms}$ the root mean square of the timing noise, and $1/\Delta \tau$ is the cadence of the observations. The ATD error then is:
\begin{equation}
    \sigma_{\Delta t} = (2\pi f \rho^2)^{-1}.
\end{equation}

\begin{figure*}[htbp]
\centering
\includegraphics[width=0.48\textwidth]{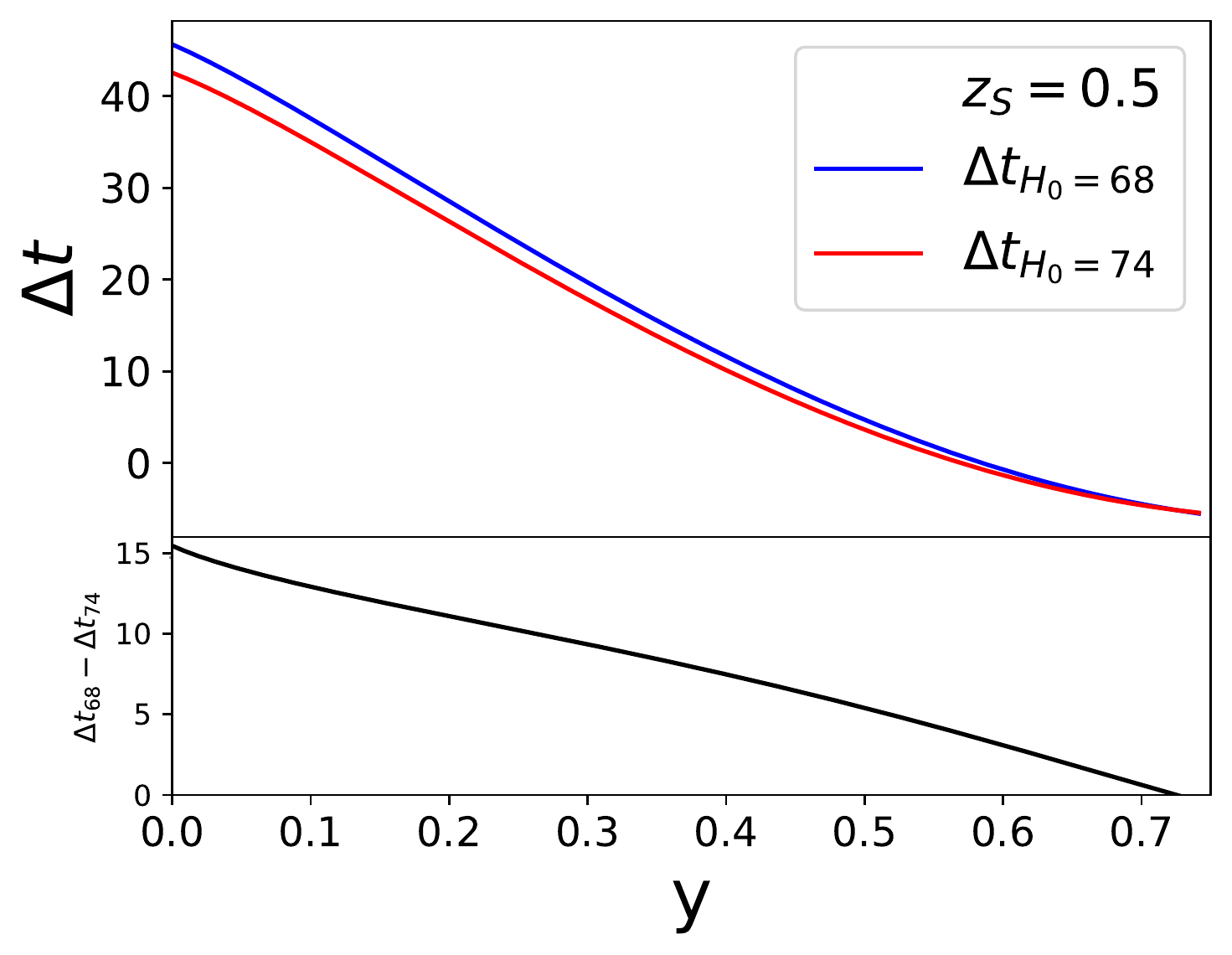}~~~
\includegraphics[width=0.48\textwidth]{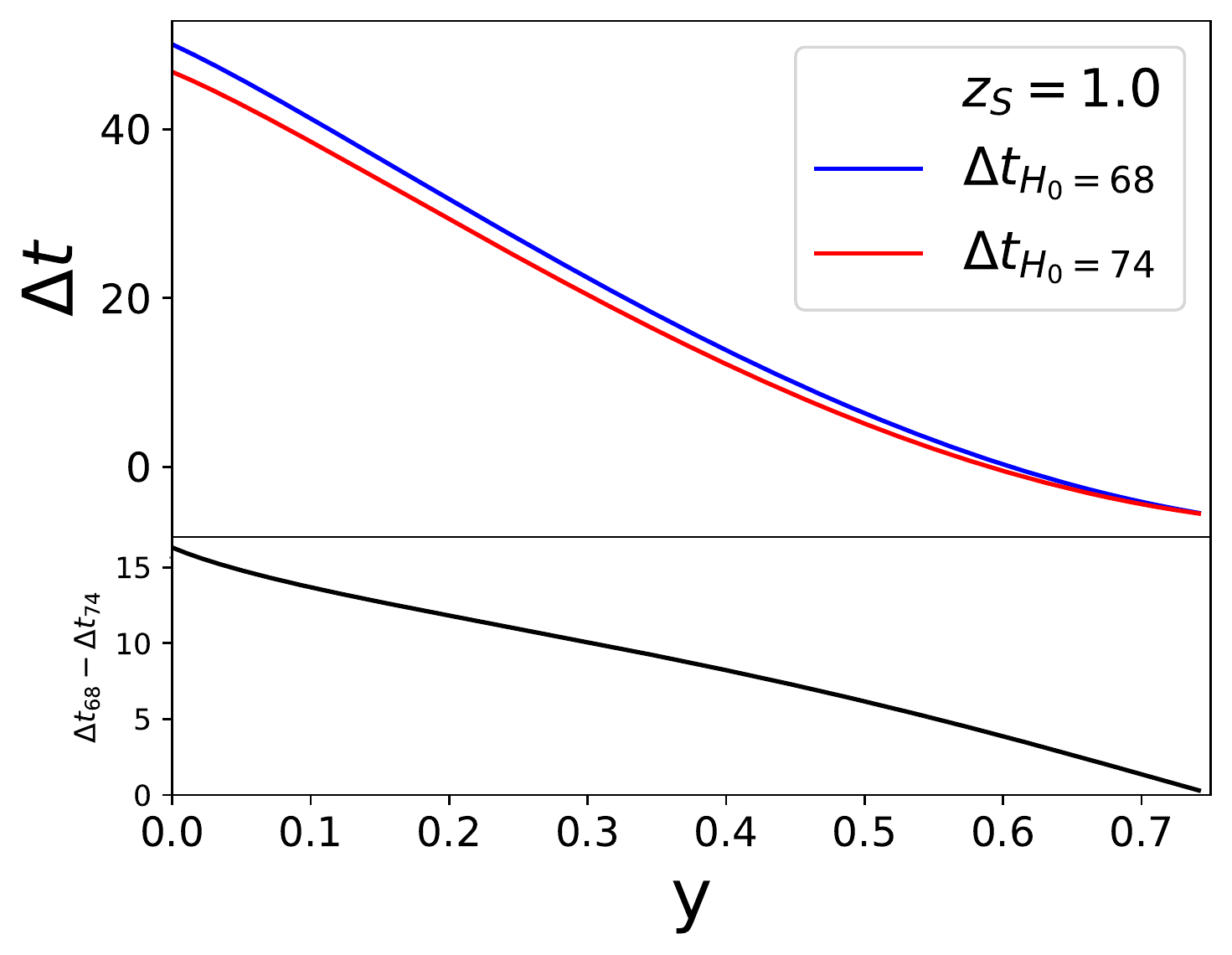}\\
~~~\\
\includegraphics[width=0.48\textwidth]{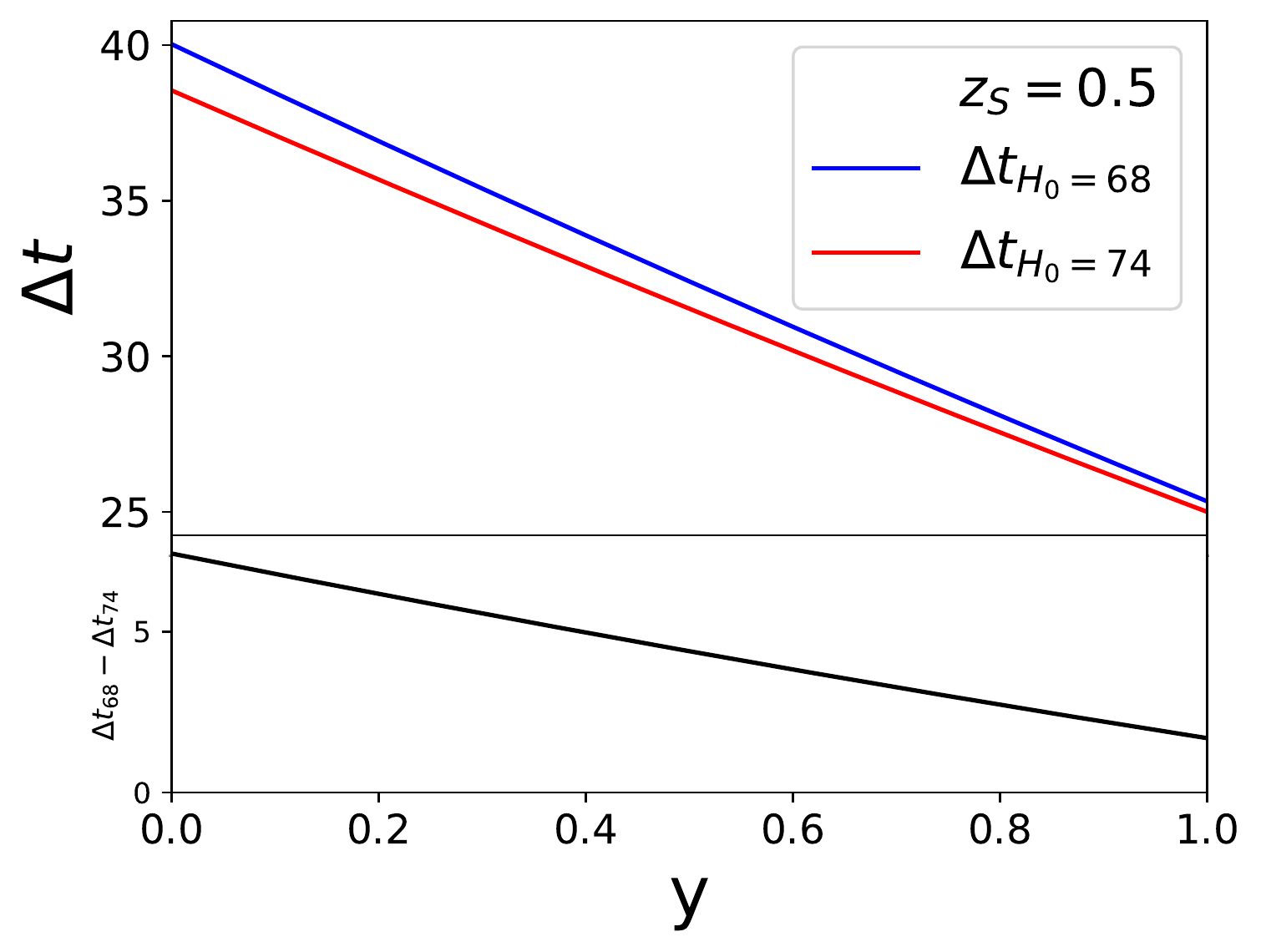}~~~
\includegraphics[width=0.48\textwidth]{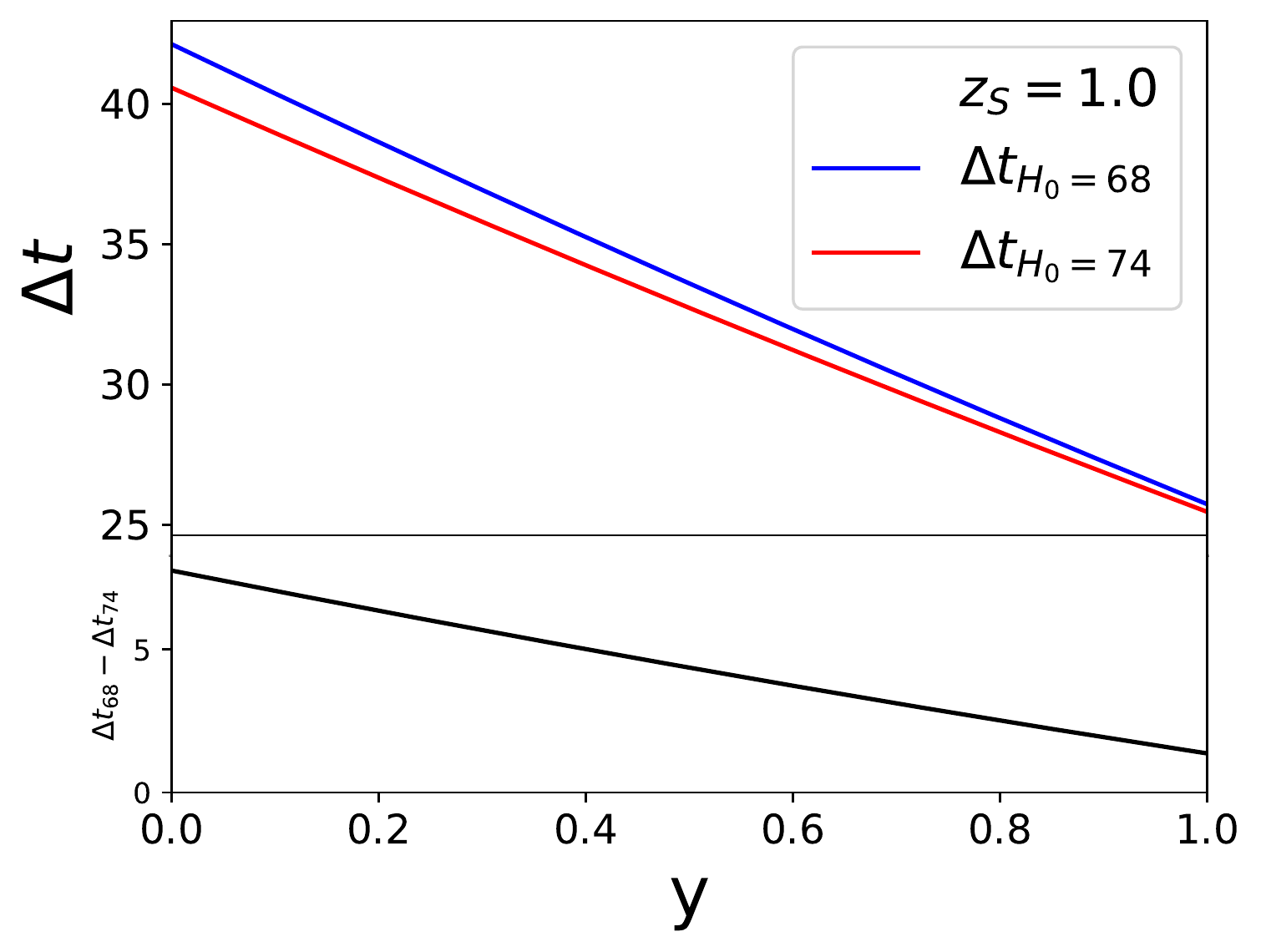}
\caption{For every figure, in the upper panel, we compare the time delays ($\Delta t$, in units of $10^5$ seconds) from different $H_0$ against the source position. $H_0=68$ is in blue and $H_0=74$ in red. In the lower panel, we show the difference between the two. Notice how the higher is the value of $y$, the smaller is the difference in the time delay. First row: NFW lens; second row: SIS lens.}
\label{plot:der_t}
\end{figure*}

\begin{figure*}[htbp]
\centering
\includegraphics[width=0.48\textwidth]{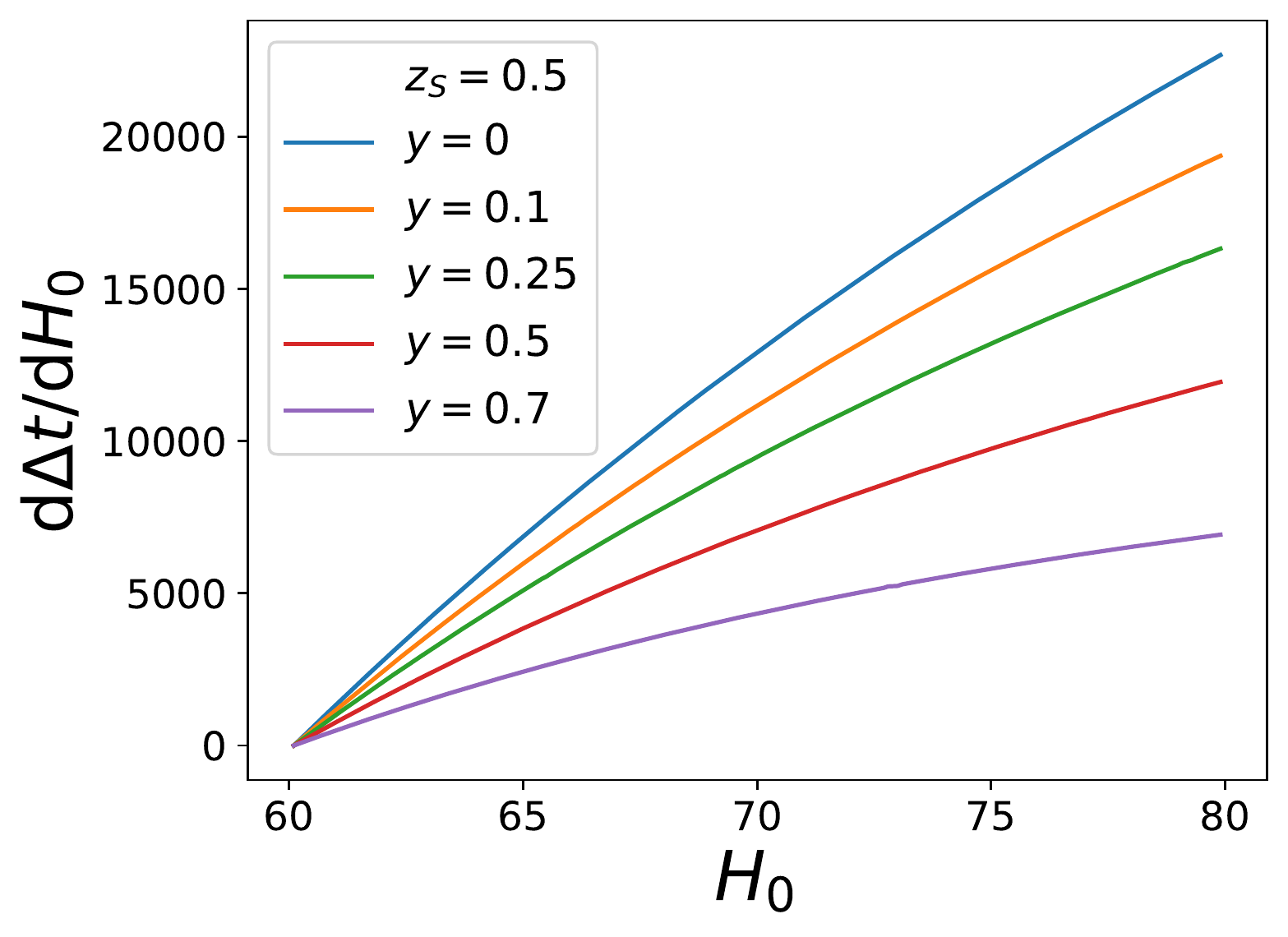}~~~
\includegraphics[width=0.48\textwidth]{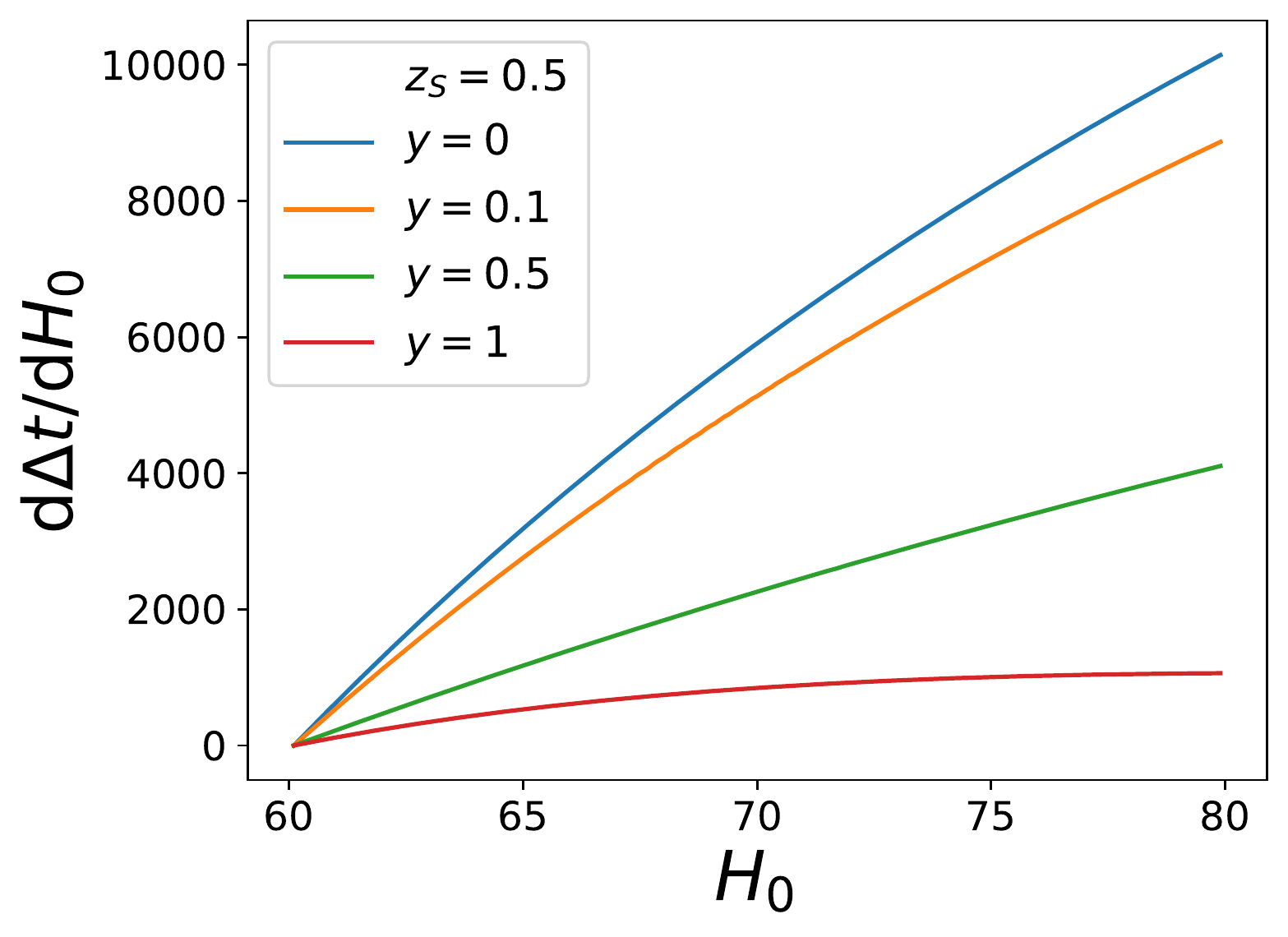}
\caption{Derivative of the arrival time difference with respect to $H_0$, for different source positions $y$. As one can see, a higher value of the derivative at low $y$, given the same difference in $\Delta t$, corresponds to a smaller difference in $H_0$ (and therefore a better constrain) w.r.t. the higher $y$ values. Left panel is for NFW lens, right one for SIS lens. In both case, the source is located at $z_S=0.5$.}
\label{plot:der_t_1}
\end{figure*}

We study two scenarios:
\begin{itemize}
 \item a state-of-the-art sample made of 65 pulsars which are nowadays known and monitored \cite{Perera:2019sca};
 \item an ``optimistic'' future sample of 1000 pulsars which might be possibly detected by SKA \cite{Bull:2018lat}.
\end{itemize}
Moreover, we also vary:
\begin{itemize}
 \item the redshift of the source, for which we consider two values, $z_S= 0.5, 1$. Instead, the lens is always placed at $z_L=0.1$;
 \item the real position of the source on the lens plane, assuming $y = 0, 0.1, 1$ for SIS and $y = 0, 0.1, 0.5$ for NFW. 
\end{itemize}
As we will see below (Tab. \ref{tab:errorsH0}), the value of the uncertainty on $H_0$ depends on $y$. A higher $y$ value means that the source is located further away (in a line-of-sight projection) from the lens. This has two main consequences. First: the lensing effect is \textit{overall} lower and weaker, thus it is intrinsically more difficult to use it for cosmological analysis. Second: the wave optics falls out of validity, and any difference rising from the geometrical vs wave optics approach fades out thus, again, any cosmological use of the time delay event is highly suppressed. For this reason, we have chosen as maximum $y$ in the NFW case the value $y=0.5$, which falls in the regime of full validity of the wave optics approach. We show this effect in \figurename~\ref{plot:der_t}. Clearly, it is shown how the arrival time difference goes to zero for high $y$. In \figurename~\ref{plot:der_t_1}, we show the dependence of the arrival time difference derivative w.r.t. to the Hubble parameter, for different values of the source position $y$. As it is clearly shown, for higher values of $y$ such profiles are ``flattened''. This means that, given the same change in the arrival time difference (i.e. given the errors $\sigma_{\Delta t}$), we will have larger variations in the error on the Hubble parameter (i.e. larger errors on $H_0$).

Other parameters are: luminosity distance $d_{\rm L}\simeq2.9$  and $6.7$ Gpc for $z_L= 0.5, 1$; for the fiducial cosmology we are assuming $H_0 = 68$ and $74$ km s$^{-1}$ Mpc$^{-1}$; the total observing time $T_{\rm obs}=10$ yr (and therefore $f_{\rm obs}=1/T_{\rm obs}=2\cdot3.17\cdot10^{-9}\text{ s}^{-1}$); the observing cadence is approximately one every 2 weeks \cite{2016IPTA}, i.e. $\Delta \tau = 2$ weeks $\simeq0.038$ yr;  and the timing noise $\sigma_{\rm rms}\approx100$ ns. Note that these values describe quite well current PTAs web of telescopes, but could be improved by future observatories. However, for what concerns the ``optimistic'' approach based on future observations, we have decided to rely on ``conservative'' values, as confirmed by recent papers (e.g. \cite{Sesana:2012, Boyle:2010rt}), thus altering only the number of observed pulsars, i.e. choosing $N_p=1000$. We summarize everything in Table~\ref{tab:sensobs}, where we also explicitly display the resulting uncertainty on the time delay, $\sigma_{\Delta t}$, expressed in days.

{\renewcommand{\tabcolsep}{1.5mm}
{\renewcommand{\arraystretch}{1.5}
\begin{table}[htbp]
\caption{Parameters used to analyze different PTAs scenarios, and corresponding error on the time arrival difference.}\label{tab:sensobs}
\centering
\begin{tabular}{ccccccc}
\hline
\hline
$z_S$ &	$\mathcal{M}$	    & $T_{obs}$ & $\sigma_{rms}$ & $\Delta \tau$ &	$N_p$  & $\sigma_{\Delta T}$ \\
	  & ($10^8 M_\odot$)	&	(yr)	& (ns)           &	(yr)      &	       & (days) \\
\hline
\hline							
\multirow{2}{*}{0.5} & \multirow{4}{*}{5} & \multirow{4}{*}{10} & \multirow{4}{*}{100} & \multirow{4}{*}{0.038} & 65& 0.835 \\
& & & & & 1000	& 0.003 \\
\multirow{2}{*}{1} & & & & & 65& 1.431  \\
& & & & & 1000	& 0.006 \\
\hline
\hline
\end{tabular}
\end{table}}}

\section{Results and discussion}\label{results}

In order to find out the precision on $H_0$ which can be achieved by the detection of a combined GW-EM gravitational lensing signal, we first calculate the expected value of the ATD in our fiducial cosmology. In the $\Lambda$CDM case we always consider two different values for the Hubble constant, i.e. $H_0 = 68$ and $74$ km s$^{-1}$ Mpc$^{-1}$, but the main results (i.e. the errors on $H_0$) are only very weakly dependent on this choice. In fact, on one hand a different value of $H_0$ changes the background history, and this results in: \textit{a)} a different possible outcome for the measured arrival time delay difference; \textit{b)} an influence on the sensitivity of the measurement to the given parameter. On the other hand, the dependence of the arrival time delay w.r.t. $H_0$ is weak, but stronger than w.r.t. to other cosmological parameters. The key point of this work is exactly to study if such (even weak) dependence can be converted into statistically useful and competitive constraints on $H_0$. Note that, despite this (even weak) dependence, the change in the uncertainty on $H_0$ is statistically negligible and not due to any flaw in our analysis but perfectly physical. In order to clarify this point even better, we have calculated the value of the error for $H_0$, assuming $H_0=71$ km s$^{-1}$ Mpc$^{-1}$, for several configurations. We got that, for example, in the case of a NFW lens, with source at $y=0$ and $z_s=0.5$, and we obtained an uncertainty on $H_0$ of $1.43$, perfectly lying in between the values given when assuming $H_0=68$ and $H_0=74$ km s$^{-1}$ Mpc$^{-1}$.
 
We have decided to show results from both values just for ``visual'' reasons: because different values of $H_0$ would give different values for the arrival time delay difference, we aim to show at least two different possible outcomes from observational measurements, and how the errors, which might be obtained from GW lensing, will automatically exclude one of the two values of $H_0$, thus concurring to alleviate the Hubble tension.

Further, we calculate the error on the ATD, $\sigma_{\Delta t}$, from each scenario we have described above, and from it we derive the possible confidence interval of an ATD detection.

Once we have our set of parameters vectors, built as $\{\Omega_m, H_0, \Delta t_{EM-GW}\}$ for the $\Lambda$CDM model and as $\{\Omega_m, w_0, H_0, \Delta t_{EM-GW}\}$ for the quiessence one, we select those values which both fit the confidence interval and cross-check the independent prior on $\Omega_m$ which is derived from \textit{Planck}.

\begin{figure*}[htbp]
\centering
\includegraphics[width=0.48\textwidth]{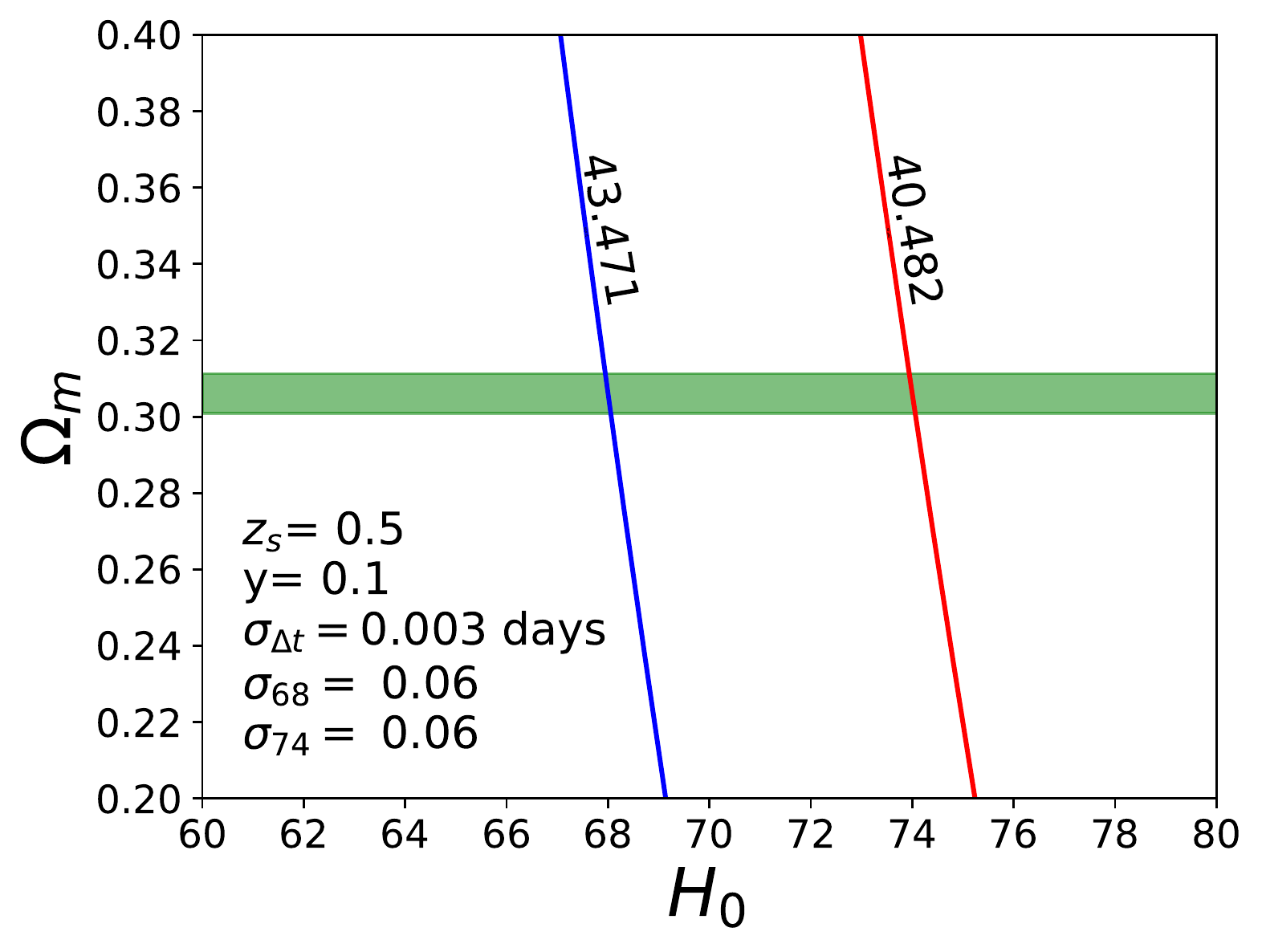}~~~
\includegraphics[width=0.48\textwidth]{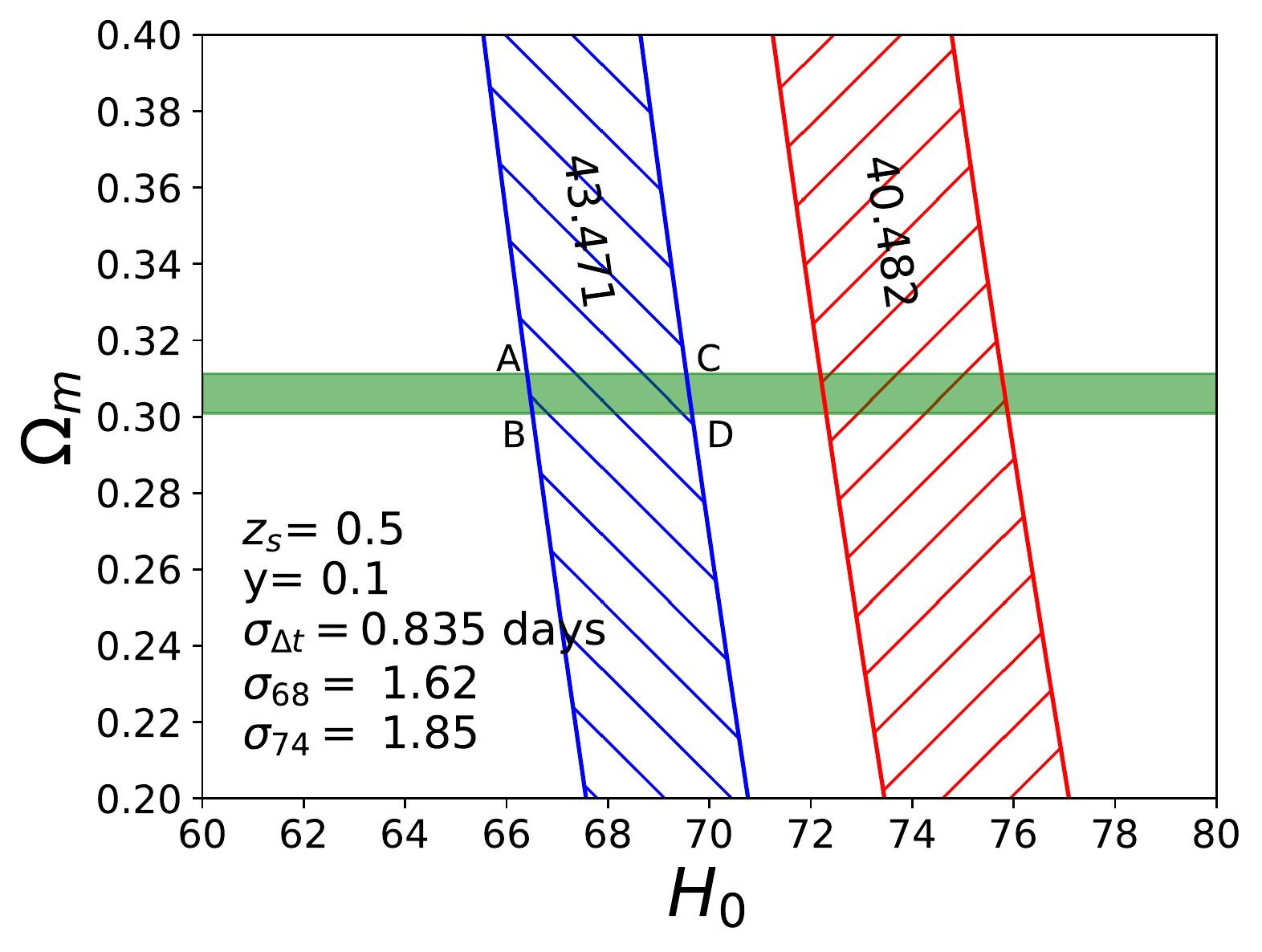}\\
~~~\\
\includegraphics[width=0.48\textwidth]{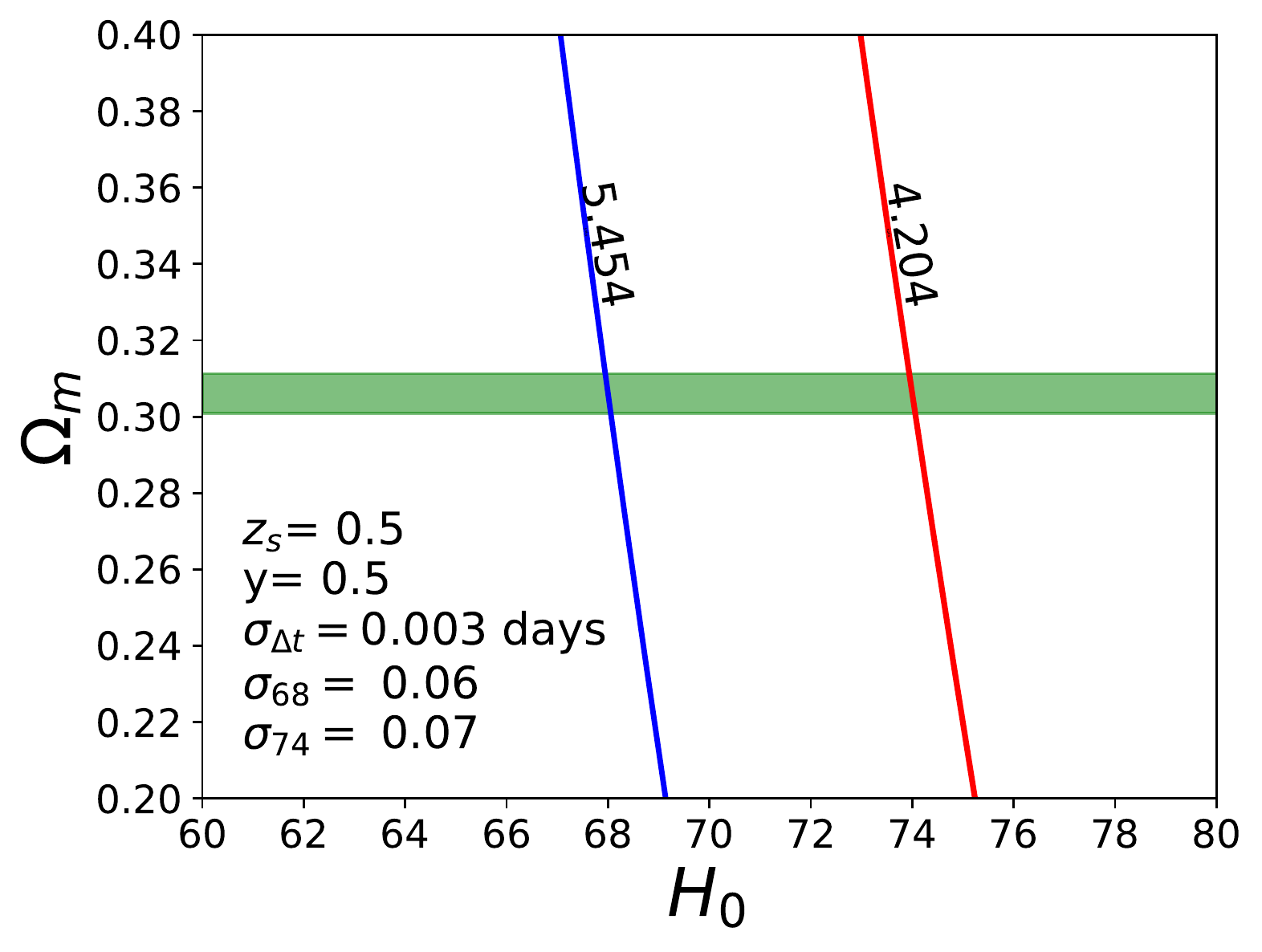}~~~
\includegraphics[width=0.48\textwidth]{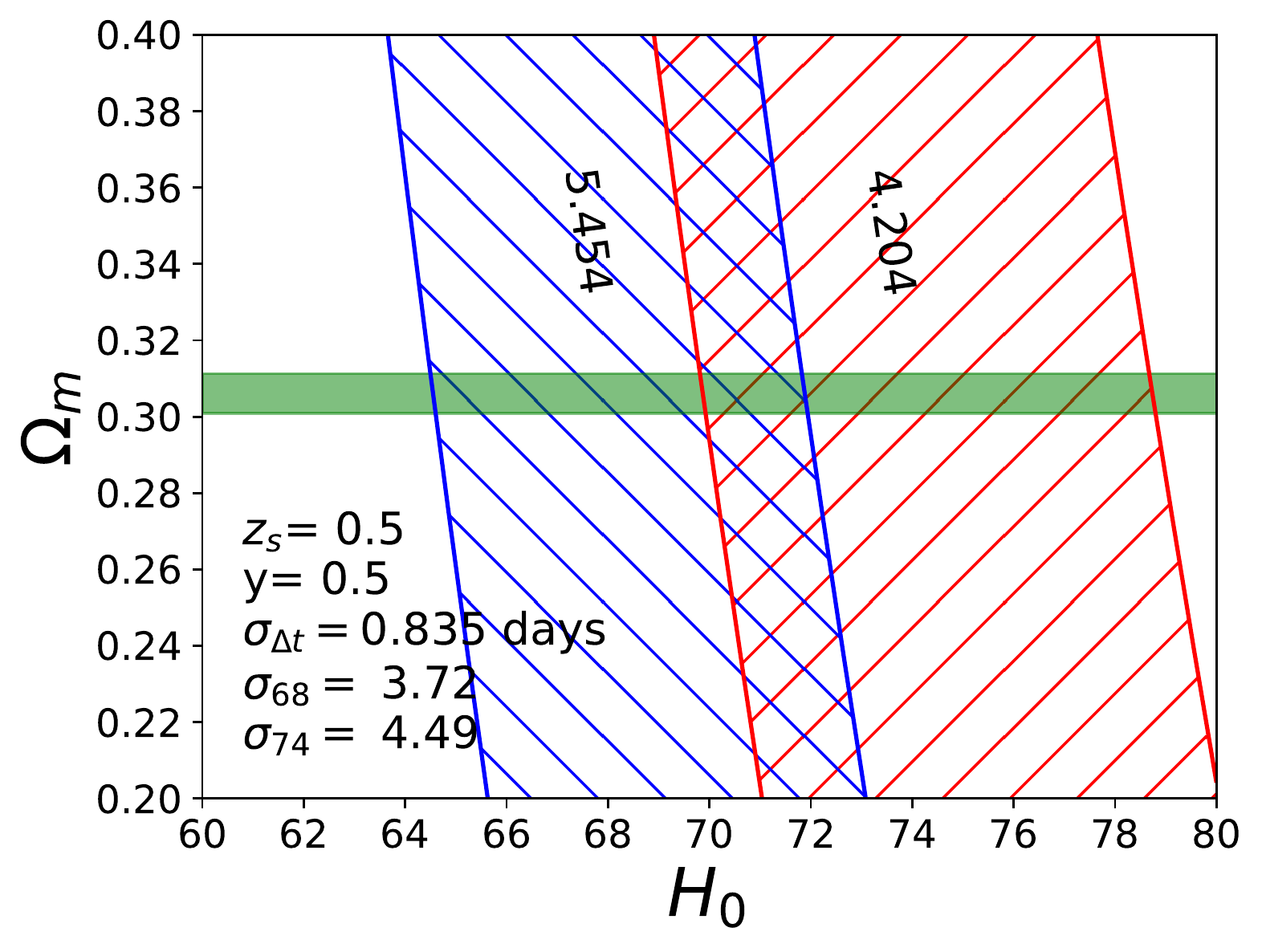}
\caption{Examples of the constraints on $H_0$ for a NFW lens. Blue hatched regions are for the case of $H_0=68$ km s$^{-1}$ Mpc$^{-1}$ and $\Omega_m=0.306$; in red for $H_0=74$ km s$^{-1}$ Mpc$^{-1}$ and same $\Omega_m$. The plots on the left column are for the future ``optimistic'' scenario, with an array of $N_p=1000$ pulsars; the right column describes the current scenario, $N_p=65$. The source here is at $z_S=0.5$. The source position $y$ is displayed in the legend of the plots. The green horizontal bar defines the prior on $\Omega_m$ from \textit{Planck}.}
\label{plot:1}
\end{figure*}

Uncertainty on $H_0$ is then calculated as shown in the first row-second column panel of Fig.~\ref{plot:1}. There, the two values of the ATD are $\Delta t = 43.471$ days (for $H_0 = 68$ km s$^{-1}$ Mpc$^{-1}$) and $\Delta t = 40.482$ days (for $H_0 = 74$ km s$^{-1}$ Mpc$^{-1}$), with an error of $\sigma_{\Delta t}= 0.835$ days, as from Table \ref{tab:sensobs}. The result of cross-checking our theoretically calculated values of the ATD with the \textit{Planck} prior is the identification of the four points $\{A, B, C, D\}$. Thus, our final estimation of the error on $H_0$ is given by the greatest distance between these points. Results of this procedure is summarized in Table \ref{tab:errorsH0} for all the cases we have taken into consideration.

\begin{figure*}[htbp]
\centering
\includegraphics[width=0.48\textwidth]{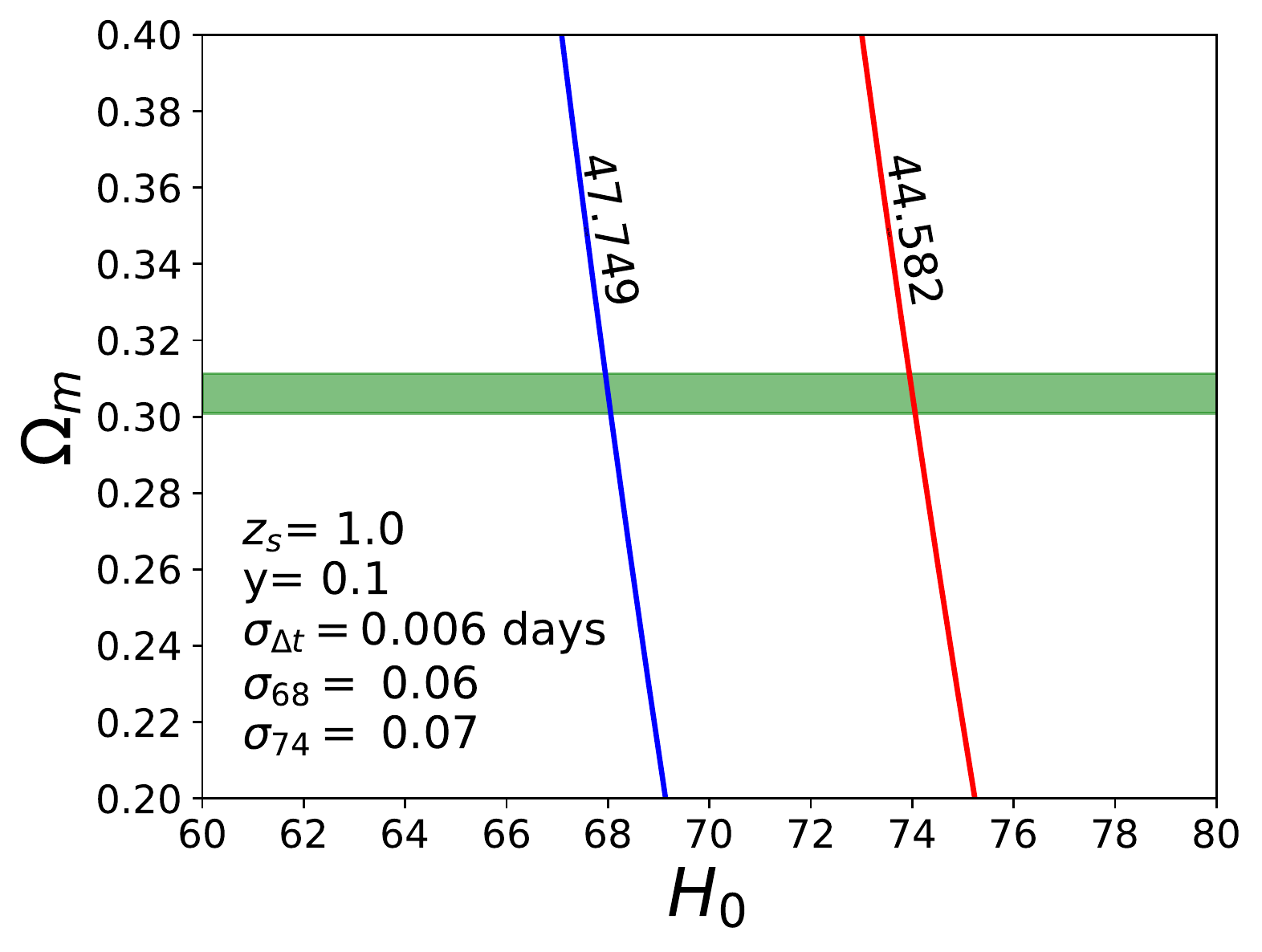}~~~
\includegraphics[width=0.48\textwidth]{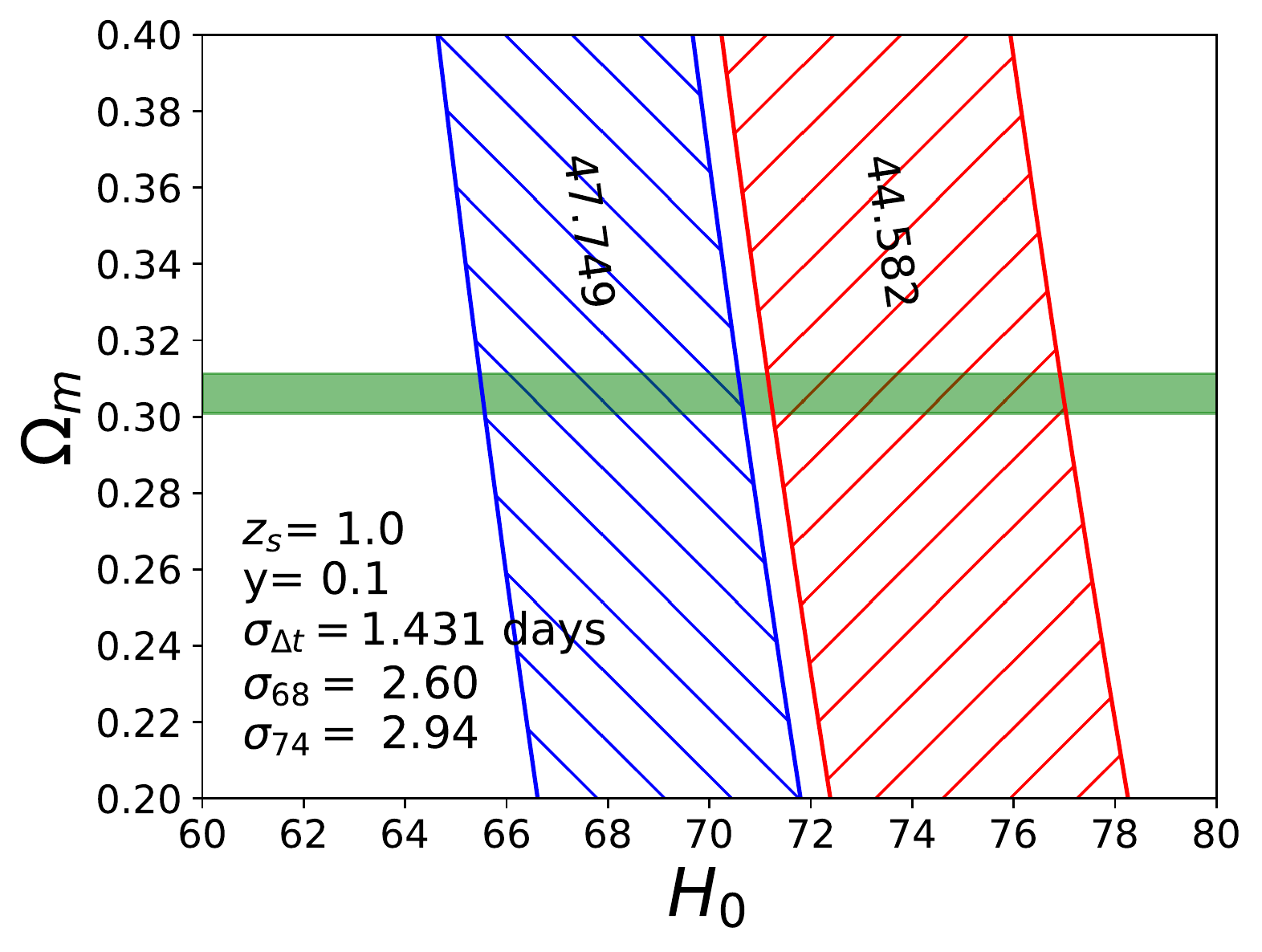}\\
~~~\\
\includegraphics[width=0.48\textwidth]{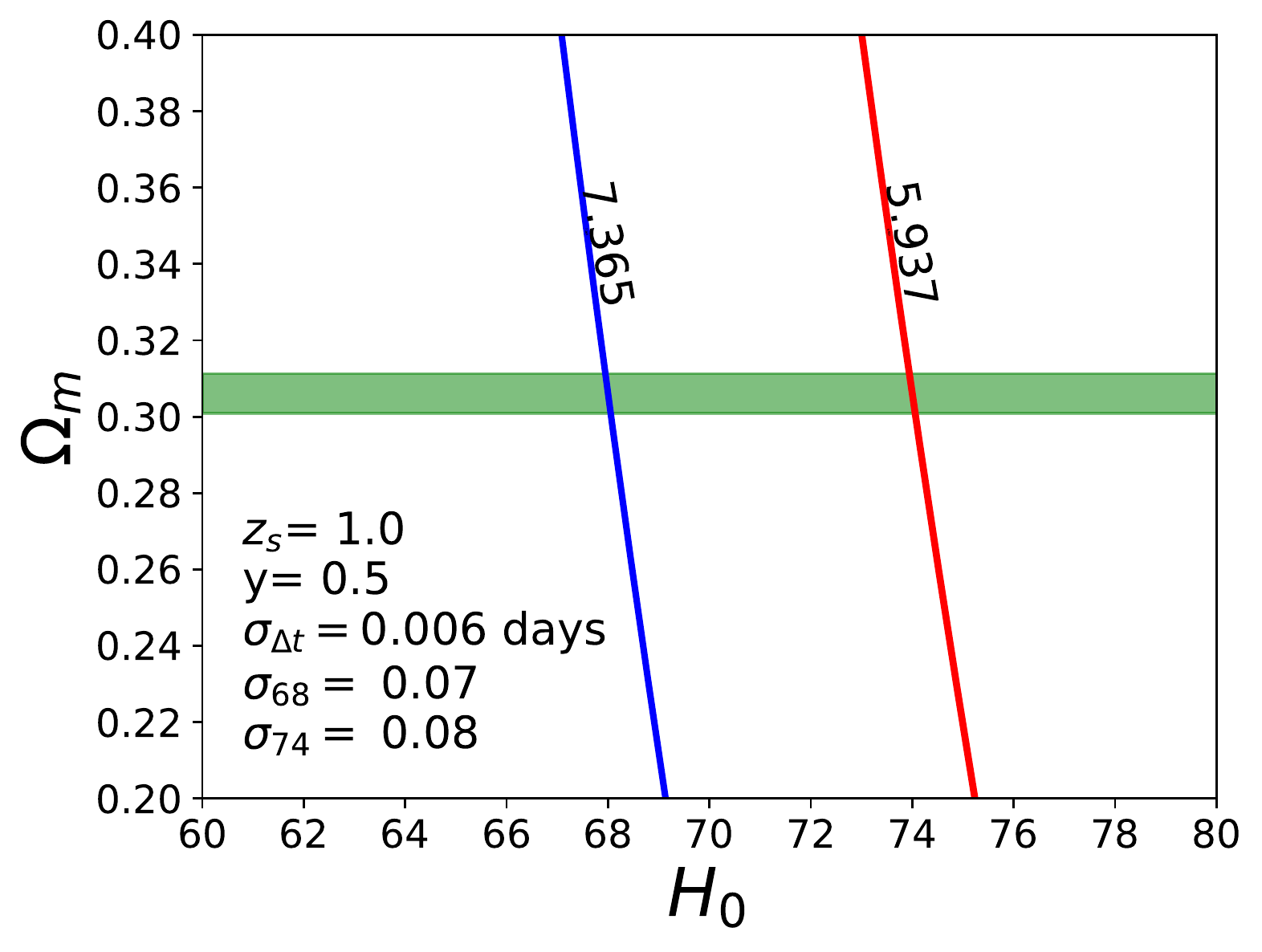}~~~
\includegraphics[width=0.48\textwidth]{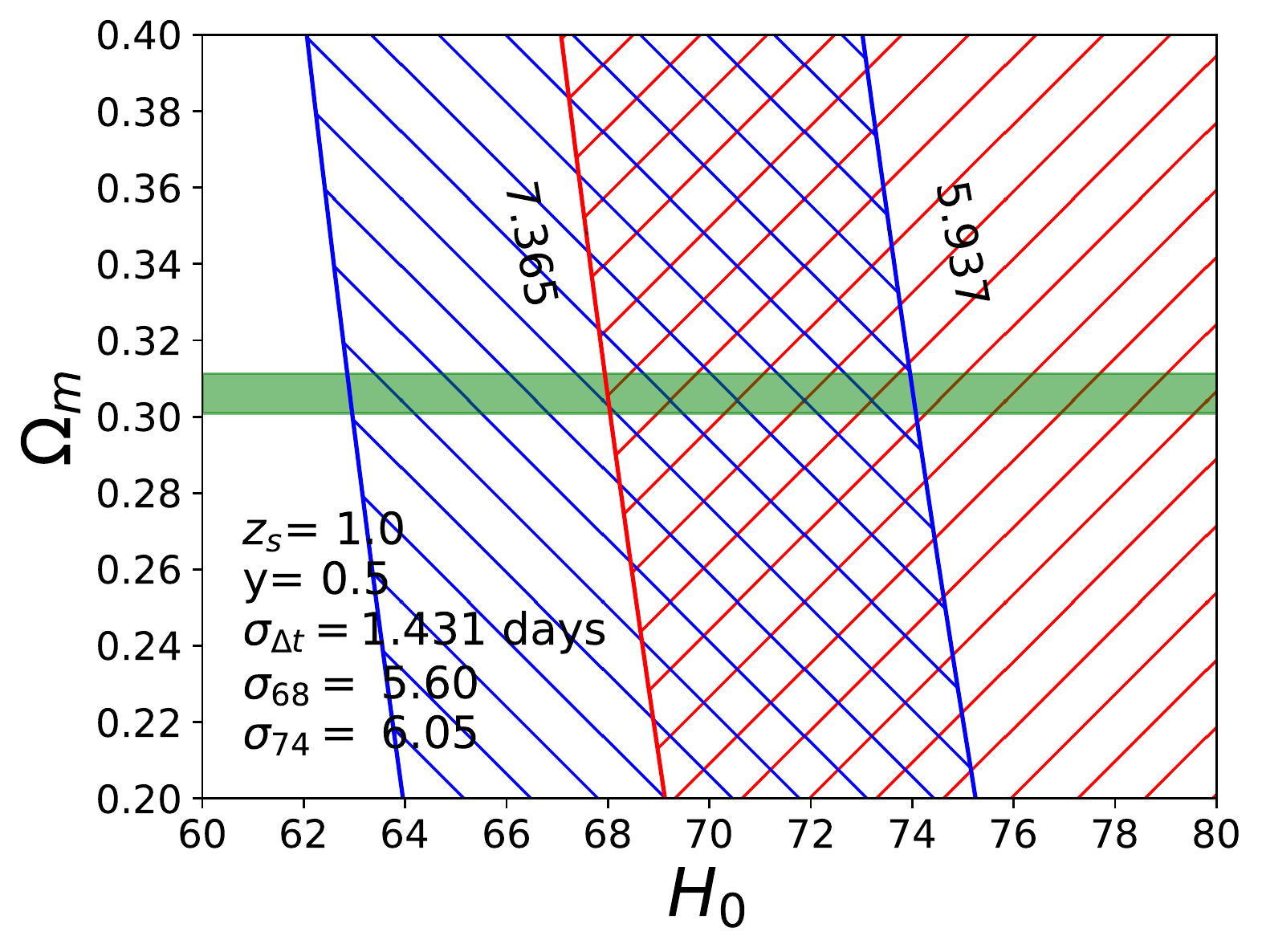}
\caption{Examples of the constraints on $H_0$ for a NFW lens. Blue hatched regions are for the case of $H_0=68$ km s$^{-1}$ Mpc$^{-1}$ and $\Omega_m=0.306$; in red for $H_0=74$ km s$^{-1}$ Mpc$^{-1}$ and same $\Omega_m$. The plots on the left column are for the future ``optimistic'' scenario, with an array of $N_p=1000$ pulsars; the right column describes the current scenario, $N_p=65$. The source here is at $z_S=1.0$. The source position $y$ is displayed in the legend of the plots. The green horizontal bar defines the prior on $\Omega_m$ from \textit{Planck}.}
\label{plot:1_1}
\end{figure*}

\begin{figure*}[ht!]
\centering
\includegraphics[width=0.48\textwidth]{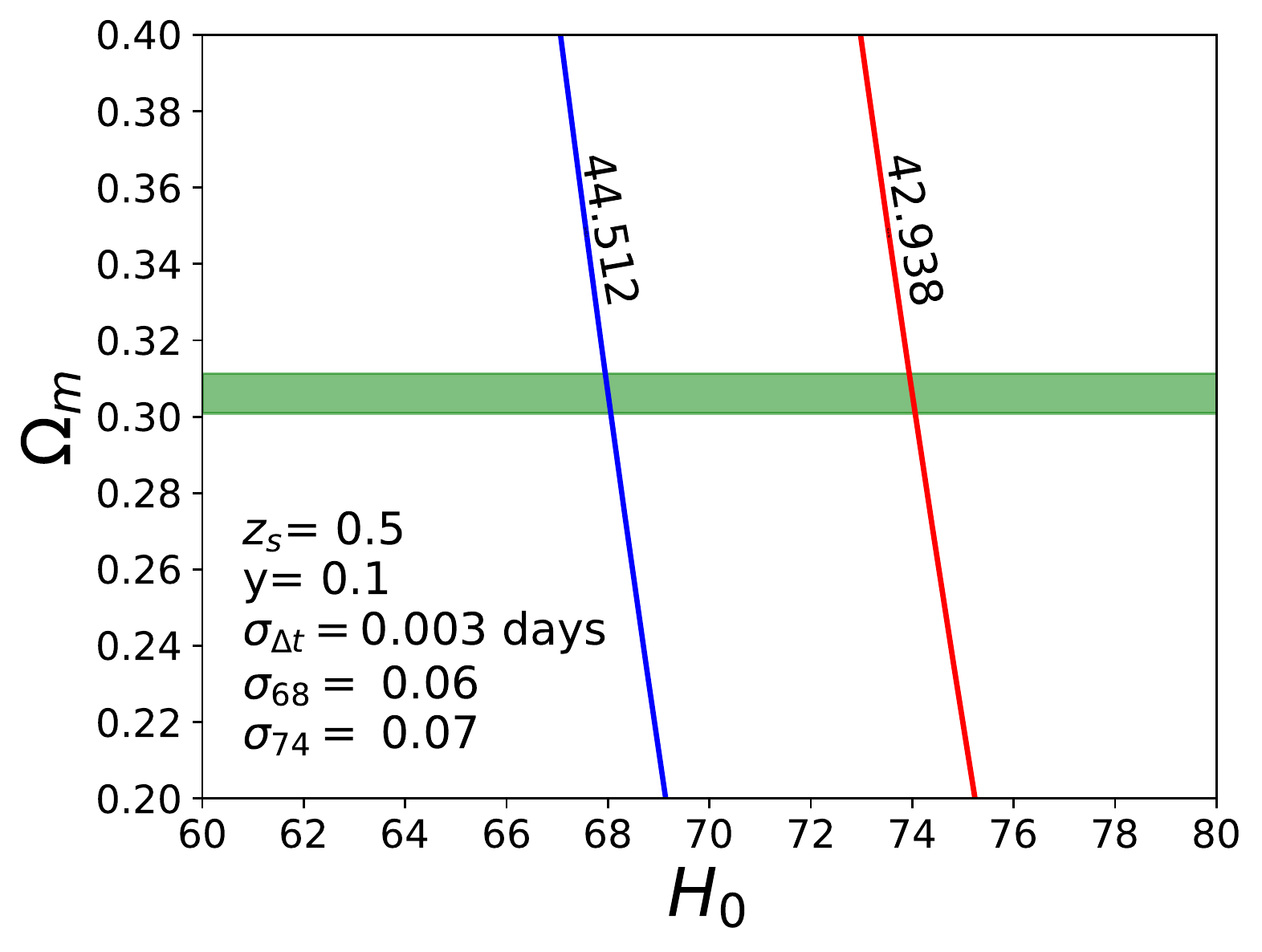}~~~
\includegraphics[width=0.48\textwidth]{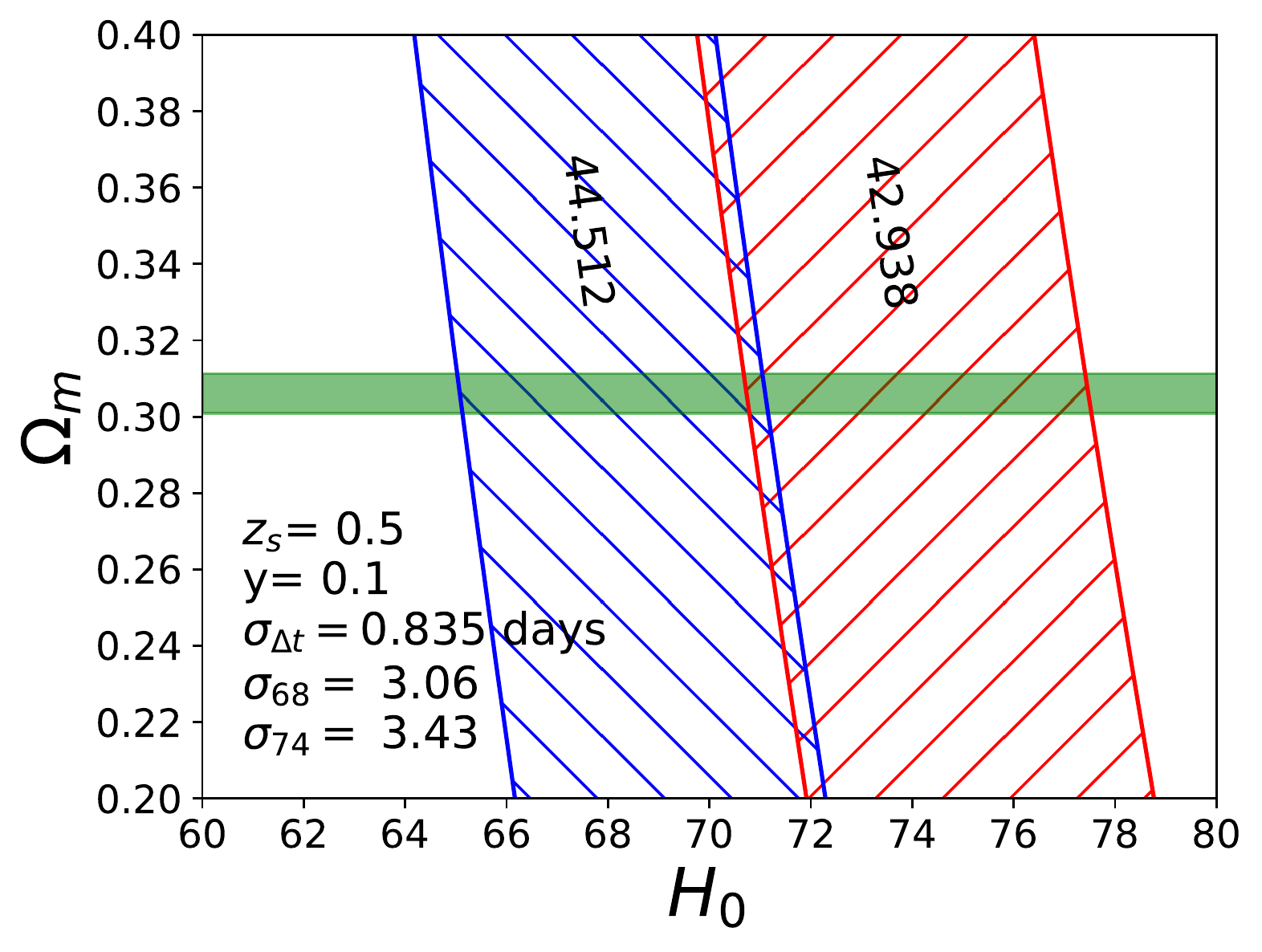}\\
~~~\\
\includegraphics[width=0.48\textwidth]{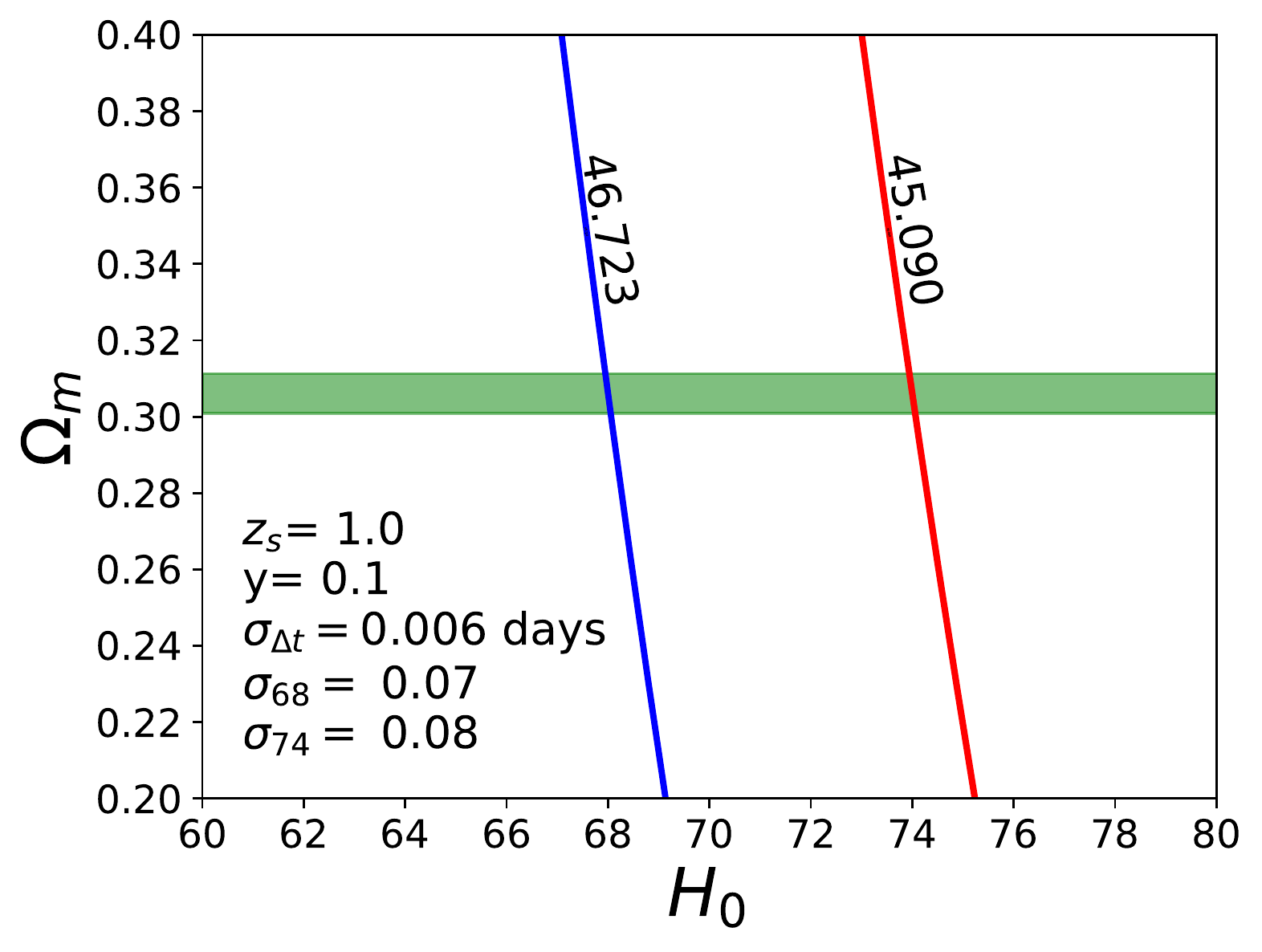}~~~
\includegraphics[width=0.48\textwidth]{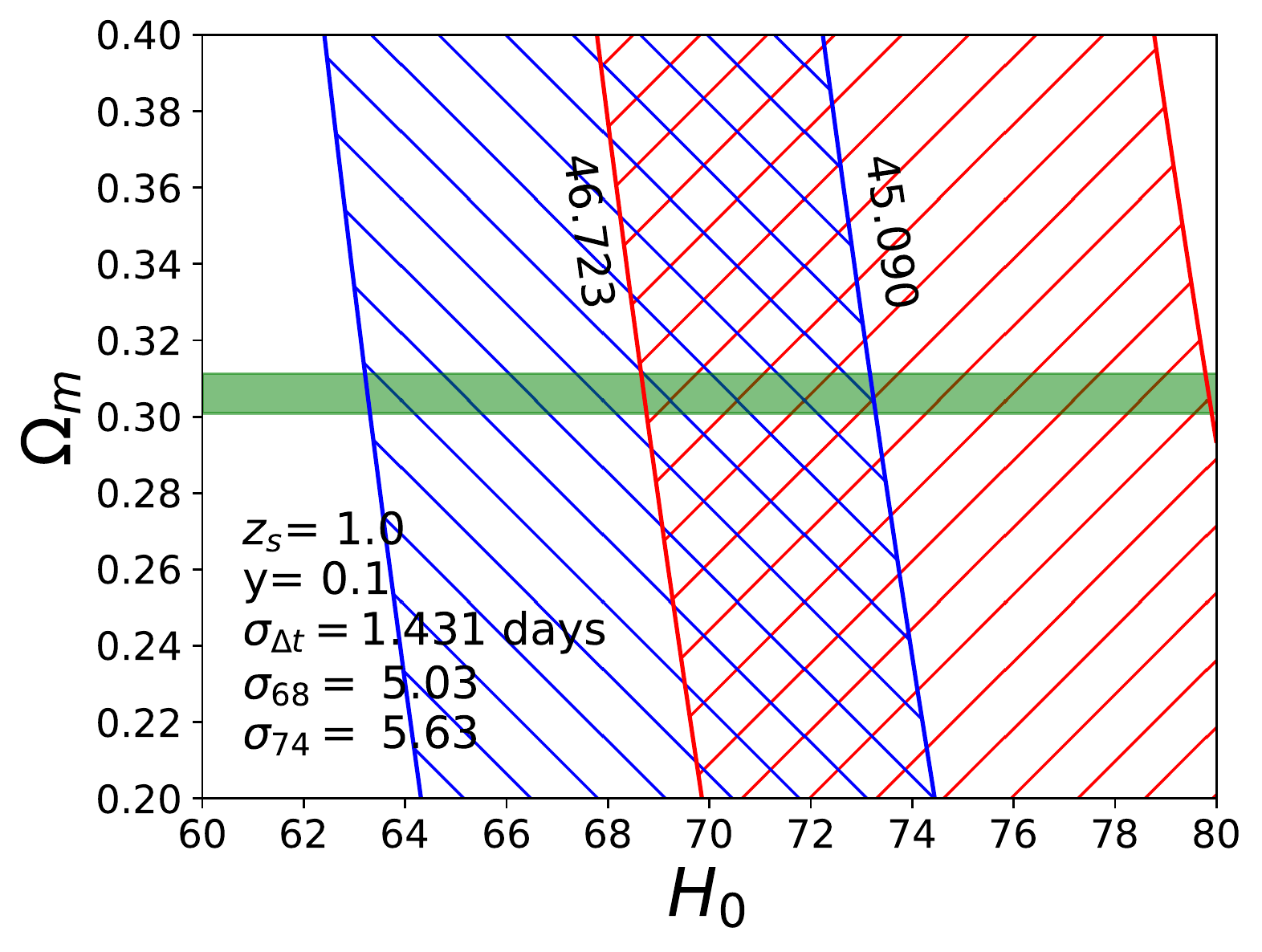}
\caption{Examples of the constraints on $H_0$ for a SIS lens. Blue hatched regions are for the case of $H_0=68$ km s$^{-1}$ Mpc$^{-1}$ and $\Omega_m=0.306$; in red for $H_0=74$ km s$^{-1}$ Mpc$^{-1}$ and same $\Omega_m$. The plots on the left column are for the future ``optimistic'' scenario, with an array of $N_p=1000$ pulsars; the right column describes the current scenario, $N_p=65$. The first row has a source at $z_S=0.5$; the second row has $z_S=1$. The source position $y$ is displayed in the legend of the plots. The green horizontal bar defines the prior on $\Omega_m$ from \textit{Planck}.}
\label{plot:2}
\end{figure*}

As we can easily spot from Table \ref{tab:errorsH0}, and as shown is some selected cases in~\cref{plot:1,plot:1_1,plot:2}, the results are quite promising.
In the figures, the hatched regions (shown just as examples of possible outcomes of future observations) correspond to the expected measured value of the ATD for our fiducial cosmological model and two different values of the Hubble constant, $H_0=68$ km s$^{-1}$ Mpc$^{-1}$ (blue) and $74$ km s$^{-1}$ Mpc$^{-1}$ (red), plus the error on the same quantity from the GW observation. The corresponding error on $H_0$ is given in the legend of the figure. The green horizontal bar defines the prior on $\Omega_m$ from \textit{Planck}. All time delays values are in days.

As we can see from Table \ref{tab:errorsH0}, when assuming a standard $\Lambda$CDM model, the estimated errors on the $H_0$ measurement in the best (``pessimistic'') \textit{present} scenario are $\sigma_{H_0}\lesssim 2.$ km s$^{-1}$ Mpc$^{-1}$ for a NFW profile (and $\lesssim 3.5$ km s$^{-1}$ Mpc$^{-1}$ for a SIS one), thus being highly competitive with the other measurements of $H_0$ available today. It is important to stress that such an error could be achieved with only one single lensing event. Thus, even if such events are supposed to be rare, they would provide fundamental information.

It is also clear that both the distance (from the observer) and the relative position (with respect to the lens) of the source have a role in the cosmological inference,. But while for the distance of the source the effect is not so strong (with the errors rising till $\sigma_{H_0}\leq 3.$ km s$^{-1}$ Mpc$^{-1}$ for NFW and $\leq 6$ km s$^{-1}$ Mpc$^{-1}$ for SIS), the relative position has a much more decisive impact, with a notable (expected) degradation of the constraints for large offsets.

For future observatories the scenario is much more positive and, we would say, decisive. In fact, uncertainty on $H_0$ might be as low as $0.1$ km s$^{-1}$ Mpc$^{-1}$ for NFW, and $0.15$ km s$^{-1}$ Mpc$^{-1}$ for SIS, even in the case of large offsets ($y=1$). Consider that the latest best measurement of the Hubble constant have both higher uncertainty ($\sigma_{H_0}=0.54$ in \cite{PlanckCollab:2018} and $1.42$ in \cite{Riess:2019cxk}), with respect to our estimations.

The quiessence case instead is different. Here, the parameters space is much bigger, the degeneracy among the cosmological parameters higher, and consequently the precision on $H_0$ is also worsened, with $\sigma_{H_0} \sim 20$ km s$^{-1}$ Mpc$^{-1}$ in the present scenario, and $\sim 14$ km s$^{-1}$ Mpc$^{-1}$ in the future one. One way to improve the error could be to have multiple observations, which will decrease the error as $1/\sqrt{n}$, with $n$ the number of observations. In that case, with, e.g., $10$ observations, we could have a much better uncertainty on $H_0$, comparable with present values from cosmic distance ladder. In this case, though, we need to consider that the probability of detection is quite low \cite{Takahashi:2016jom} and therefore it will be hard to collect an high number of observations.

{\renewcommand{\tabcolsep}{1.5mm}
{\renewcommand{\arraystretch}{1.8}
\begin{table*}[htbp]
\caption{Uncertainty on the Hubble constant, calculated in different observational scenarios, as described in Sec.~\ref{Methodology}.} \label{tab:errorsH0}
\centering
\begin{tabular}{cc|c|c|c|c|c|c|c|c}
\hline									
\hline	
\multicolumn{2}{c||}{$z_S$}	&	\multicolumn{4}{c|}{0.5}	&	\multicolumn{4}{c}{1}	\\										
\hline																	
\multicolumn{2}{c||}{\multirow{2}{*}{$\sigma_{\Delta T}$ (days)}}	&	\multicolumn{2}{c|}{\multirow{2}{*}{0.835}} &	\multicolumn{2}{c|}{\multirow{2}{*}{0.003}} &	\multicolumn{2}{c|}{\multirow{2}{*}{1.431}} &	\multicolumn{2}{c}{\multirow{2}{*}{0.006}}	\\
\multicolumn{2}{c||}{} & \multicolumn{2}{c|}{} & \multicolumn{2}{c|}{} & \multicolumn{2}{c|}{} & \multicolumn{2}{c}{} \\
\hline
\multicolumn{2}{c||}{$H_0$ (km s$^{-1}$ Mpc$^{-1}$)} &	68	&	74	&	68	&	74	&	68	&	74	&	68	&	74	\\
\hline															
\multicolumn{2}{c||}{$y~\downarrow$} & \multicolumn{8}{c}{NFW - $\Lambda$CDM}	\\	
\hline																
\multicolumn{2}{c||}{0} & 1.37 & 1.55 & 0.06 & 0.06 & 2.19 & 2.47 & 0.06 & 0.07 \\
\multicolumn{2}{c||}{0.1} & 1.62 & 1.85 & 0.06 & 0.06 & 2.60 & 2.94 & 0.06 & 0.07 \\
\multicolumn{2}{c||}{0.5} & 3.72 & 4.49 & 0.06 & 0.07 & 5.60 & 6.05 & 0.07 & 0.08 \\
\hline
\multicolumn{2}{c||}{}		&	\multicolumn{8}{c}{NFW - quiessence}														\\
\hline																		
\multicolumn{2}{c||}{0} & 14.50 & 15.80 & 12.10 & 14.20 & 14.70 & 16.20 & 12.60 & 13.70 \\
\multicolumn{2}{c||}{0.1} & 14.60 & 16.00 & 12.60 & 14.20 & 15.00 & 16.70 & 12.60 & 13.70 \\
\multicolumn{2}{c||}{0.5} & 16.50 & 18.20 & 13.10 & 14.30 & 16.90 & 19.30 & 12.70 & 13.90 \\
\hline																	
\multicolumn{2}{c||}{}		&	\multicolumn{8}{c}{SIS - $\sigma_\ast=220$ (km/s) - $\Lambda$CDM}			\\				
\hline	
\multicolumn{2}{c||}{0} & 2.80 & 3.15 & 0.06 & 0.07 & 4.56 & 5.13 & 0.07 & 0.08 \\
\multicolumn{2}{c||}{0.1} & 3.06 & 3.43 & 0.06 & 0.07 & 5.03 & 5.63 & 0.07 & 0.08 \\
\multicolumn{2}{c||}{1} & $>$10 & 9.70 & 0.10 & 0.10 & $>$10 & $>$10 & 0.17 & 0.16 \\
\hline																	
\multicolumn{2}{c||}{}		& \multicolumn{8}{c}{SIS - $\sigma_\ast = 220$ km/s - quiessence}		\\							\hline
\multicolumn{2}{c||}{0} & 15.80 & 17.30 & 12.90 & 14.20 & 16.10 & 18.70 & 12.70 & 13.90 \\
\multicolumn{2}{c||}{0.1} & 16.00 & 17.60 & 12.90 & 14.20 & 16.40 & 19.00 & 12.70 & 13.90 \\
\multicolumn{2}{c||}{1} & $>$20.00 & $>$20.00 & 13.20 & 14.40 & $>$20.00 & $>$20.00 & 12.80 & 13.90 \\	
\hline
\hline
\end{tabular}
\end{table*}}}

\section{Conclusions}
\label{conclusions}

In this work we have studied the possibility to constrain the Hubble constant by gravitational lensing from a multi-messenger detection of both GW and EM radiation of the same event. We have shown that in a $\Lambda$CDM universe, with one single event of lensing detected from a pulsar population similar to the present PTA sample (i.e. counting $\sim65$ pulsars), and applying a prior on $\Omega_m$ from \textit{Planck}, we could match current precision on $H_0$ as obtained from distance ladder methods \cite{Riess:2019cxk}. If we consider future observatories, like SKA, which will have a much larger array of pulsars ($\sim1000$), the errors will be improved by two orders of magnitudes, giving a considerable contribution in clarifying the Hubble tension problem. In the case of relaxing the cosmological background assumptions, for example by considering a quiessence model, the errors on $H_0$ are much bigger, as expected, given the larger degeneracy between the cosmological parameters. In this case, only multiple observations would help to refine the uncertainty to comparable and competitive levels, but the rarity of GW lensing would make this goal much harder.

In a series of forthcoming papers, we are going to analyze in deeper details the role of the lens modelling in the cosmological background determination and ``if and how'' the ATD could be useful to provide high accuracy constraints also on the astrophysical scale. In fact, we are aware that the mass sheet degeneracy (MSD) is important when one extrapolates the lens parameters \textit{only} from lensing data, as recently remarked in \cite{Kochanek:2019ruu}. But actually the same results from  H0LiCOW seem to show that such $10\%$ barrier can be overcome, if one takes into account a variety of many independent types of observations related to the lens. Both \cite{Chen:2019,Blum:2020mgu} suggest that \textit{``stellar velocity data do resolve the MSD''}, and H0LiCOW itself makes an intensive and fully-comprehensive use of multiple (dedicated) data and theoretical mass models which alleviate this issue.

In this paper, though, we have not focused on this aspect because our main and primary goal was to show that GW lensing is feasible and can be very effective in constraining $H_0$, maybe assuming some ``ideal'' constraints on the mass model or that, at least, they are not dominant w.r.t. the GW signal. This aspect is not so crucial at the present time, but will be fundamental in our future optimistic scenario, where the errors from GW lensing should/would be (most probably) under-dominant. We want to stress again here that the most important ingredient in this recipe is not simply the GW lensing itself, but the wave optics approach to the GW lensing event. This is the key, what makes the arrival time difference detectable in the GW waveform and usable for cosmological purposes. In the next future, we are going to show how to face this problem and how, actually, the same GW lensing can be used as a further independent tool to constrain the mass model of the lens (provided that other independent data are given).

\bibliographystyle{apsrev4-1}
\bibliography{biblio}{}

\end{document}